\documentclass{aa}

\usepackage[varg]{txfonts}
\usepackage{graphicx}
\usepackage{natbib}             %for A&A : reimplements the LaTeX \cite command 
\usepackage{threeparttable}      %to follow the A&A style
\usepackage{booktabs} % To thicken table lines
\usepackage[colorlinks=true]{hyperref}
\hypersetup{citecolor=blue}
\hypersetup{linkcolor=red}

\bibpunct{(}{)}{;}{a}{}{,}           %to follow the A&A style

\newcommand{\kms}{\mbox{${\rm km\,s}^{-1}$}}

\begin{document}

\title{Hot Exoplanet Atmospheres Resolved with Transit Spectroscopy
(HEARTS)\thanks{Based on observations made at ESO 3.6m telescope at the La Silla Observatory under ESO programme 098.C-0304 (PI: Ehrenreich).}}
\subtitle{VII. Detection of sodium on the long-transiting inflated sub-Saturn KELT-11\,b}
\author{D. Mounzer\inst{1}
\and C. Lovis\inst{1}
\and J. V. Seidel\inst{2,1}
\and O. Attia\inst{1}
\and R. Allart\inst{3}
\and V. Bourrier\inst{1}
\and D. Ehrenreich\inst{1}
\and A. Wyttenbach\inst{1}
\and N. Astudillo-Defru\inst{4}
\and T. G. Beatty\inst{5}
\and H. Cegla\inst{6,7}
\and K. Heng\inst{8,6,9}
\and B. Lavie\inst{1}
\and M. Lendl\inst{1}
\and C. Melo\inst{10}
\and F. Pepe\inst{1}
\and J. Pepper\inst{11}
\and J. E. Rodriguez\inst{12}
\and D. Ségransan\inst{1}
\and S. Udry\inst{1}
\and E. Linder\inst{13}
\and S. Sousa \inst{14}
}
\institute{Geneva Observatory, University of Geneva, Chemin Pegasi 51b, CH-1290 Versoix,  Switzerland \\
 \email{dany.mounzer@unige.ch}
\and European Southern Observatory, Alonso de Córdova 3107, Vitacura, Santiago, Chile
\and Department of Physics and Institute for Research on Exoplanets, Université de Montréal, Montreal, QC, Canada
\and Departamento de Matemática y Física Aplicadas, Universidad Católica de la Santísima
Concepción, Concepción, Chile
\and Department of Astronomy and Steward Observatory, University of Arizona, Tucson, AZ 85721
\and Department of Physics, Astronomy \& Astrophysics Group, University of Warwick, Coventry CV4 7AL, United Kingdom
\and Centre for Exoplanets and Habitability, University of Warwick, Coventry, CV4 7AL, United Kingdom
\and University of Bern, Center for Space and Habitability, Gesellschaftsstrasse 6, 3012 Bern, Switzerland
\and Ludwig Maximilian University, University Observatory Munich, Scheinerstrasse 1, Munich 81679, Germany
\and Portuguese Space Agency,  Estrada das Laranjeiras, n .º 205, RC, 1649-018,  Lisboa,  Portugal
\and Department of Physics, Lehigh University, 16 Memorial Drive East, Bethlehem, PA 18015, USA
\and Center for Data Intensive and Time Domain Astronomy, Department of Physics and Astronomy, Michigan State University, East Lansing, MI 48824, USA
\and Lucerne School of Engineering and Architecture, Technikumstrasse 21, 6048 Horw, Switzerland
\and Instituto de Astrofísica e Ciências do Espaço, Universidade do Porto, CAUP, Rua das Estrelas, 4150-762 Porto, Portugal
}
\date{Received: May 10, 2022 / Accepted: August 19, 2022}

%ABSTRACT
\abstract{High-resolution transmission spectroscopy has allowed for in-depth information on the composition and structure of exoplanetary atmospheres to be garnered in the last few years,  especially in the visible and in the near-infrared. Many atomic and molecular species have been detected thanks to data gathered from state-of-the-art spectrographs installed on large ground-based telescopes.  Nevertheless, the Earth daily cycle has been limiting observations to exoplanets with the shortest transits.}
{The inflated sub-Saturn KELT-11\,b has a hot atmosphere and orbits a bright evolved subgiant star, making it a prime choice for atmospheric characterization. The challenge lies in its transit duration -- of more than seven hours -- which can only be covered partially or without enough out-of-transit baselines when observed from the ground.}
{To overcome this constraint, we observed KELT-11\,b with the HARPS spectrograph in series of three consecutive nights, each focusing on a different phase of the planetary orbit: before, during, and after the transit. This allowed us to gather plenty of out-of-transit baseline spectra, which was critical to build a spectrum of the unocculted star with sufficient precision. Telluric absorption lines were corrected using the atmospheric transmission code \textsc{Molecfit}. Individual high-resolution transmission spectra were merged to obtain a high signal-to-noise transmission spectrum to search for sodium in KELT-11\,b's atmosphere through the $\sim$ 5900 \AA ~doublet.}
{Our results highlight the potential for independent observations of a long-transiting planet over consecutive nights. Our study reveals a sodium excess absorption of 0.28 $\pm$ 0.05 \% and 0.50 $\pm$ 0.06 \% in the Na D1 and D2 lines, respectively. This corresponds to 1.44 and 1.69 times the white-light planet radius in the line cores.  Wind pattern modeling tends to prefer day-to-night side winds with no vertical winds, which is surprising considering the planet bloatedness. The modeling of the Rossiter-Mclaughlin effect yields a significantly misaligned orbit, with a projected spin-orbit angle of $\lambda$ = -77.86$^{+2.36}_{-2.26}{}^\circ$.}
{Belonging to the under-studied group of inflated sub-Saturns, the characteristics of KELT-11\,b -- notably its extreme scale height and long transit -- make it an ideal and unique target for next-generation telescopes. Our results as well as recent findings from HST, TESS, and CHEOPS observations could make KELT-11\,b a benchmark exoplanet in atmospheric characterization.}

\titlerunning{Na detection on KELT-11\,b}
\maketitle

%Observation Log
\begin{table*}[h]
\caption{HARPS observations' log of KELT-11.}
\label{table:log}
\centering
\begin{tabular}{c c c c c c c c}
\toprule
Epoch \# & Night \# & Date & Spectra & $T_\mathrm{exp} [s]$ & Airmass \tablefootmark{a} & Seeing \tablefootmark{a} & SN@550nm \tablefootmark{a} \\
\toprule
Epoch 1 & Night 1 & 2017-02-01 & 37 (3 in, 34 out)\tablefootmark{b} & 300 & 1.05-1.45 & 0.53-0.72 & 98-114 \\ \midrule
& Night 2 & 2017-02-14 & 28 (28 out) & 400 & 1-1.3 & 0.5-0.9 & 100-127 \\
Epoch 2 & Night 3 & 2017-02-15 & 69 (13 out, 51 in, 5 out) & 400 & 1-2.3 & 0.8-1.7 & 55-135 \\
& Night 4 & 2017-02-16 & 28 (28 out) & 400 & 1.05-1.8 & 0.6-0.9 & 75-110 \\ \midrule
& Night 5 & 2017-03-05 & 43 (43 out) & 300 & 1.05-2 & 0.45-0.9 & 75-110 \\
Epoch 3 & Night 6 & 2017-03-06 & 93 (12 out, 68 in, 13 out) & 300 & 1.05-2.3 & 0.6-1.1 & 55-100 \\
& Night 7 & 2017-03-07 & 46 (46 out) & 300 & 1.05-1.6 & 0.45-1.1 & 85-110 \\ \bottomrule
\end{tabular}
\tablefoot{Night 1, 3, and 6 are transit nights. For each transit, their day-before and day-after observations are divided into epochs. \\
\tablefoottext{a}{The value of the left corresponds to the minimum, and the one on the right corresponds to the maximum.}
\tablefoottext{b}{Night 1 has too few in-transit spectra and was thus not used in the final transmission spectrum.}}
\end{table*}

% INTRODUCTION
\section{Introduction}

After more than a quarter century of exoplanetary discoveries, the field of atmospheric characterization has expanded significantly. Nowadays, with most discovered exoplanets found with transit photometry, one of the key tools that allowed for the growth of the characterization field has been transit spectroscopy. The observation of transiting systems with spectrographs allows us to derive the transmission spectrum \citep{SS2000}, which contains a plethora of information on the physical properties of the planet's atmosphere.  

The first planet detected by transit HD\,209\,458\,b \citep{Charb2000, Henry2000, Mazeh2000} is the most studied exoplanet and also the first one in which the detection of an atomic signature, through the Na D doublet at 589 nm, was claimed, with the use of the STIS spectrograph on the Hubble Space Telescope (HST) \citep{Charb2002}. This sodium feature in the visible band is one of the easiest to detect in the upper part of the hot atmosphere of giant exoplanets and one of the most detected species \citep{Madhu2019}. HD\,209\,458\,b's sodium detection was recently challenged as \citet{CB2020, CB2021} suggest that the observed signatures are a combination of the Rossiter-McLaughlin (RM) effect (\citealt{Ross1924}; \citealt{McL1924}) and center-to-limb variations (CLVs; \citealt{Czesla2015}; \citealt{Yan2017}), as their transit observations using HARPS-N and CARMENES spectrographs show no absorption when taking these effects into account.

While the first observations of exoplanet atmospheres were carried out from space at medium spectral resolution $\mathcal{R} \equiv \lambda / \Delta \lambda \sim  5~500$, ground-based measurements with high-resolution spectrographs followed. This resulted in the first detection of absorption in the sodium doublet from the ground by \citet{Red2008}, on another well-studied exoplanet HD\,189\,733\,b using the High Resolution Spectrograph  ($\mathcal{R}  \sim  60~000$) and the potential detection of the sodium signature on HD\,209\,458\,b from the ground by \citet{Snell2008} with the High Dispersion Spectrograph ($\mathcal{R} \sim  60~000$). Both used various techniques to bypass telluric contamination's systematic effects.

Since then, transit spectroscopy has allowed for the detection of other atomic and molecular traces such as carbon, oxygen \citep{VM2004}, hydrogen (e.g., \citealt{Ehren2015}), water vapor (e.g., \citealt{Dem2013}), titanium oxyde \citep{Sed2017}, helium (e.g., \citealt{Allart2018}). These detections have helped to understand and discover atmospheric processes such as circulation (e.g., \citealt{Snell2010}), evaporation (e.g., \citealt{Lecav2010}), temperature gradients (e.g., \citealt{Huit2012}), and clouds and hazes (e.g., \citealt{Sing2016}).

We report in this article our investigation and detection of sodium absorption in the upper layers of KELT-11\,b's atmosphere. KELT-11\,b lies on the edge of the Neptune desert \citep{Mazeh2016} and is amongst the most inflated exoplanets known, being larger than Jupiter and less massive than Saturn. This result is based on two sets of 3-day observations from the High-Accuracy Radial-velocity Planet Searcher (HARPS) echelle spectrograph \citep[][]{Mayor2003} on the ESO 3.6m telescope in La Silla, Chile ($\mathcal{R}  \sim  115~000$), for the Hot Atmospheres Resolved with Transit Spectroscopy (HEARTS) survey under ESO program 098.C-0304 (PI: Ehrenreich). This survey has already achieved several sodium detections through transit spectroscopy \citep{Wytt2017, Seidel2019, Hoeij2020, Seidel2020b}.

%OBSERVATIONS AND DATA REDUCTION
\section{Observations and data reduction}
\label{section:2}
\subsection{3-day observation method}
Transits of close-in planets typically last for a few hours, so that during a single night the full transit duration can be observed together with enough out-of-transit spectra to perform transmission spectroscopy. However, this has in turn biased detections of sodium absorption from the ground to exoplanets with short transit durations. (less than five hours, such as \citet{Borsa2021} on WASP-121b; \citet{McCloat2021} on KELT-10; or \citet{Allart2020} on WASP-127b). KELT-11\,b transits in front of its host star for 7.1 hours.  Such long transits are usually observed with space telescopes -- which has been done recently on KELT-11 by \citet{Colon2020} with HST, SPITZER and TESS observations -- as it is difficult to get a sufficient out-of-transit baseline on a single night from the ground. To bypass this issue, we observed the system in sets of three nights in a row; before, during, and after the transit. This ensures that we record enough out-of-transit spectra in order to build a precise out-of-transit master spectrum (master-out) with high signal-to-noise ratio (S/N), which is paramount in order to follow the method used in \citet{Wytt2015}. Having a long baseline also helps getting an effective telluric correction. We further discuss the benefits, limitations and viability of this method in Sect. \hyperlink{section.4}{4} and \hyperlink{section.5}{5}.

% Harps observations
\subsection{HARPS observations of KELT-11}
KELT-11\,b orbits one of the brightest stars with a transiting planet (V = 8.0) every P = 4.736 days (\citealt{Beatty2017}; \citealt{Pepp2017}). It is an exceptionally inflated sub-Saturn-mass exoplanet (scale height H = 2763 km), with radius of 1.35 $\pm$ 0.10 $R_J$ and mass of 0.171 $\pm$ 0.015 $M_J$, which lead to a low surface gravity ($log\,g_\mathrm{p}$ = 2.407$^{+0.080}_{-0.086}$). It has a high equilibrium temperature ($T_\mathrm{eq}$ = 1712$^{+51}_{-46}$ K) due to the proximity to its star.

We observed three transits of KELT-11\,b over seven nights using the HARPS echelle spectrograph. The HARPS CCD detector is divided into two parts, which can be distinguished by the spectra it produces: a blue part from 3800 to 5300 $\AA$~and a red part from 5380 to 6900 $\AA$. All the KELT-11 data is summarized in Table \hyperlink{table.1}{1}. The first night of observations contains a partial transit, without having out-of-transit baseline observations from the previous and next night. The two other transits, on the other hand, follow this observational strategy of 3-day measurements.

The HARPS observations were reduced with the HARPS Data Reduction Software (DRS; version 2.2.5). The software extracts the 72 spectral orders one by one; then, the spectra are flat-fielded using calibrations made before the observations, deblazed and calibrated in wavelength. This results in 1D spectra which range from 3800 \AA \,to 6900 \AA \,with a varying step depending on the wavelength. The wavelength grid remains the same for all spectra. The sets of three nights in a row use the calibration frames of their first night instead of changing the calibrations every night.

We observed transits of KELT-11\,b on 1 February 2017, 15 February 2017 and 5 March 2017 and recorded 344 spectra of 300 or 400 seconds exposure time, which are all used in our analysis as they do not exhibit any anomalies. As shown in Table \hyperlink{table.1}{1}, each set of observations is separated in epochs, as the nights of out-of-transit observations are used in combination with the in-transit nights.

%DACE
\subsection{System parameters refinement} \label{subsection:dace}
No sign of sodium was reported in \citet{Zak2019} using the same KELT-11 dataset as in this study, where orbital parameters from \citet{Pepp2017} were used to determine which spectra are in- or out-of-transit. We consider a spectrum as in-transit as long as the mid-exposure time fits in one of the transits calculated from $T_C$, $P$ and $T_\mathrm{23}$ (Table \ref{table:params}). Several sets of observations from ground-based and space telescopes (SPITZER, HST, TESS and CHEOPS) were obtained since then \citep{Beatty2017, Colon2020, Benz2021} and improved the orbital parameters of KELT-11\,b. These parameters change the configuration of in- and out-of-transit spectra substantially, in particular the difference of mid-transit time is approximately one hour and twenty minutes. This causes a misclassification of 12 to 16 spectra per transit night. We may expect that the outdated ephemeris had an impact on the results. We obtained a total of 122 in-transit spectra and 222 out-of-transit spectra (Table \ref{table:log}). 

In similar fashion to \citet{Allart2019, Bourrier2020}, we refined the orbit parameters using the DACE platform\footnote{Data \& Analysis Center for Exoplanets (DACE),  see \url{https://dace.unige.ch}} by analyzing the radial velocity data from HARPS, APF and HIRES, using the transit parameters from \citet{Beatty2017} and \citet{Colon2020} as Gaussian priors. We assign to each instrument a quadratically additive error which is applied on each data point. The priors used are detailed in Table \ref{table:a1} in Appendix \ref{appendix:DACE}. We fix the eccentricity to zero as to assume a circular orbit.

All data points are fitted to a Keplerian model \citep{Del2016} combined with activity detrending \citep{Del2018}. The fit is performed using a Markov chain Monte Carlo (MCMC) algorithm \citep{Diaz2014,Diaz2016}, shown in Fig \ref{fig:DACE} in Appendix \ref{appendix:DACE}. As a sanity check, we performed this orbit refinement with nonzero eccentricity and found no significant discrepancy compared to the circular model. The physical and orbital parameters derived from this procedure and from previous papers are summarized in Table \hyperlink{table.2}{2}.

% Parameters fitted table
\begin{table}[t]
\caption{Physical and orbital parameters of the KELT-11 system}
\label{table:params}
\small
\centering
\begin{tabular}{l l c}
\toprule \toprule
Parameter & Symbol  [Unit] & Value \\
\toprule
\multicolumn{3}{c}{-- \textit{Stellar Parameters} --} \\
Stellar mass & $M_\mathrm{\star}$ [$M_\mathrm{\odot}$] & 1.44 $\pm$ 0.07\tablefootmark{a}\\
Stellar radius & $R_\mathrm{\star}$ [$R_\mathrm{\odot}$] & 2.69 $\pm$ 0.04\tablefootmark{a}\\ 
Effective temperature & $T_\mathrm{eff}$ [K] & 5375 $\pm$ 25\tablefootmark{a}\\
Metallicity & [Fe/H] [dex] & 0.17 $\pm$ 0.07\tablefootmark{a} \\
Surface gravity & $log\,g_\mathrm{\star}$ [cgs] &  3.7 $\pm$ 0.1\tablefootmark{a}\\
%Projected rotation velocity & v~sin~i$_\star$[km s$^{-1}$] & 2.66 $\pm$ 0.50\tablefootmark{b}\\
\multicolumn{3}{c}{-- \textit{Planetary Parameters} --} \\
Planet mass & $M_\mathrm{p}$ [$M_\mathrm{J}$] & 0.205 $\pm$ 0.017\tablefootmark{d} \\
Planet radius & $R_\mathrm{p}$ [$R_\mathrm{J}$] & 1.35 $\pm$ 0.10\tablefootmark{a}\\
Equilibrium temperature & $T_\mathrm{eq}$ [K] & 1712$^{+51}_{-46}$ \tablefootmark{b} \\
Surface gravity & log $g_\mathrm{p}$ [cgs] & 2.407$^{+0.080}_{-0.086}$ \tablefootmark{b} \\
\multicolumn{3}{c}{-- \textit{System Parameters} --} \\
Transit epoch & $T_\mathrm{c}$ [$BJD_\mathrm{TDB}$]& 2455498.95898 \tablefootmark{d} \\
& & $\pm$ 0.00178 \\
Orbital period & P [d] & 4.73620865 \tablefootmark{d}  \\
& & $\pm$ 0.00000377 \\
Transit duration (P$_1$ to P$_4$) & $T_\mathrm{14}$ [h] & 7.10 $\pm$ 0.02 \tablefootmark{c} \\
Transit duration (P$_2$ to P$_3$) & $T_\mathrm{23}$ [h] & 6.25 $\pm$ 0.02 \tablefootmark{c} \\
Planet-to-star radius ratio & $R_\mathrm{p}$ / $R_\mathrm{\star}$ & 0.0519 $\pm$ 0.0026 \tablefootmark{d}\\
Orbital semi-major axis & a [AU] & 0.06230 $\pm$ 0.00104 \tablefootmark{d} \\
Orbit inclination & i [deg] & 85.3 $\pm$ 0.02 \tablefootmark{d} \\
Eccentricity & e & 0 (fixed) \tablefootmark{d}\\
Stellar RV semi-amplitude & $K_\mathrm{\star}$ [m s$^{-1}$] & 19.49 $\pm$ 1.51 \tablefootmark{d}\\
Planet RV semi-amplitude & $K_\mathrm{p}$ [km s$^{-1}$] & 143 $\pm$ 11 \tablefootmark{e}\\
Systemic velocity & $\gamma$ [km s$^{-1}$] & 35.1033 $\pm$ 0.0003 \tablefootmark{e}\\
\multicolumn{3}{c}{-- \textit{Rossiter-McLaughlin Parameters} --} \\
Limb-darkening coefficient & $u_\mathrm{1~Sloan~g}$ & 0.695 $\pm$ 0.017 \tablefootmark{b}\\
Limb-darkening coefficient & $u_\mathrm{2~Sloan~g}$ & 0.117 $\pm$ 0.013 \tablefootmark{b}\\
Projected spin-orbit angle & $\lambda$ [deg] & -77.86$^{+2.36}_{-2.26}$ \, \tablefootmark{f}\\
Projected rotation velocity & $v sin i_\mathrm{\star}$ [km s$^{-1}$]& 1.99$^{+0.06}_{-0.07}$ \, \tablefootmark{f}\\
\bottomrule
\end{tabular}
\tablefoot{
\tablefoottext{a}{\citet{Beatty2017}}
\tablefoottext{b}{\citet{Pepp2017}}
\tablefoottext{c}{\citet{Colon2020}} 
\tablefoottext{d}{This work (DACE fit with other papers' priors)}
\tablefoottext{e}{This work (fitted separately)}
\tablefoottext{f}{This work (Rossiter-McLaughlin analysis, see Sect. \ref{section:RM})}
}
\end{table}

% Molecfit
\subsection{Telluric correction using \textsc{Molecfit}}
\label{sect:molecfit}
The removal of telluric features has been an important step in transmission spectroscopy when using ground-based telescopes, as information can be hidden or biased due to the molecules in Earth's atmosphere. Water and oxygen contamination is especially prevalent and can strongly affect spectral lines. In our case, telluric water lines around the Na D doublet  (between 5850 \AA ~and 6000 \AA) reach $\sim$ 10\% in depth on average (up to 25\% for a single spectrum), which is nonnegligible, while oxygen lines reach $\sim$ 50\% in depth. As seen in \citet{Allart2017}, water lines also vary greatly with airmass. 

To correct these telluric lines, we used ESO's package \textsc{Molecfit} \citep[][version 4.0.0]{Smette2015, Kausch2015}. This tool creates an ultra high resolution telluric spectrum ($\mathcal{R}  \sim  4\,000\,000$) of the molecular species chosen -- H$_2$O and O$_2$ in our case -- based on a line-by-line radiative transfer model. This model is built from an atmospheric profile combined with a Global Data Assimilation System (GDAS) weather profile provided by the National Oceanic and Atmospheric Administration (NOAA). The atmospheric profile is established using the Reference Forward Model \citep{Rem2001} which accounts for pressure, temperature, humidity and abundance of molecular species as a function of altitude, time, location and airmass. The merged atmospheric profile can be characterized either by a fixed grid or a natural grid. The former has a fixed number of layers (50) describing the variation of parameters from 0 to 120 km, while the latter is more accurate as it uses up to 150 layers. This natural grid, while heavier to compute, is more sensitive and allows a better telluric correction.

\begin{figure}[t]
\resizebox{\hsize}{!}{\includegraphics{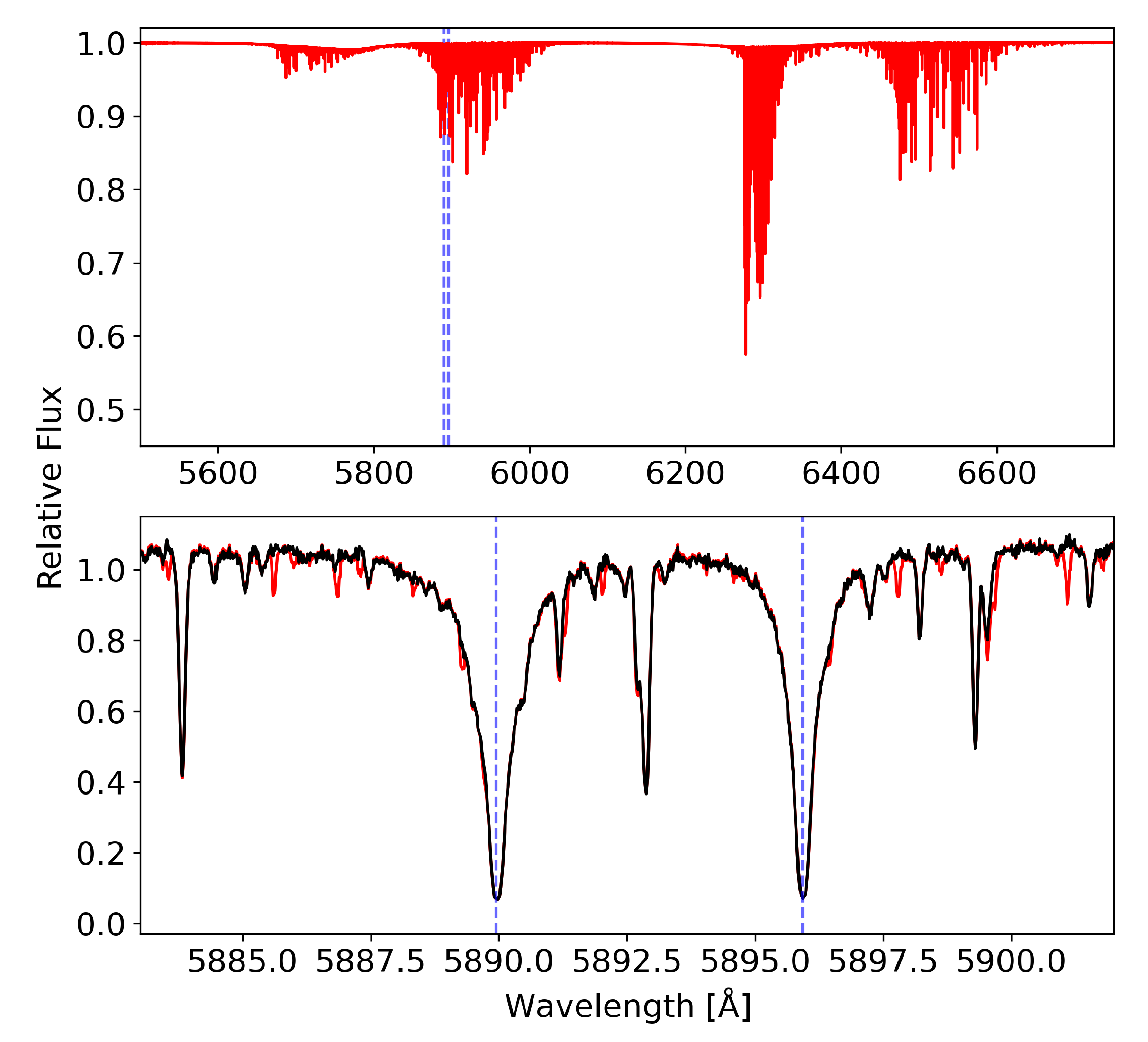}}
\caption{Influence of telluric contamination by H$_2$O and O$_2$ molecules on a spectrum of KELT-11 from HARPS observations (red CCD). We note that the resulting total correction is shown, as individual spectrum correction varies greatly case-by-case. \textbf{Top}: Applied telluric correction obtained using \textsc{Molecfit}. \textbf{Bottom}: Zoomed-in telluric correction around Na doublet ($\sim$ 5890 \AA), \textit{black} being the corrected spectrum and \textit{red} the noncorrected one.}
\label{fig:TAC}
\end{figure}

\textsc{Molecfit} was first used on HARPS spectra by \citet{Allart2017} in the search for water vapor on the atmosphere of HD\,189\,733\,b. We follow the article guidelines on telluric contamination removal with similar inital parameters (Table \ref{table:mol1}) and fitted wavelength ranges (Table \ref{table:mol2}) summarized in Appendix \ref{appendix:mol}.  This correction method has proven effective in recent works using high-resolution spectrographs such as \citet{Seidel2019, Allart2019, Allart2020, Cabot2020, Tab2021} and in telluric correction reviews \citep{Lang2021,Ulmer2019}. We tested out the influence of most parameters in order to optimize the correction: these matter less than the choice of the fitted regions. These wavelength bins must be carefully chosen following several conditions: they should contain at least one telluric absorption line which must be the main component(s); they should be surrounded with enough continuum to allow for an optimal fit of the line; and finally no stellar feature should be of similar depth as the telluric line(s).  The telluric correction we obtained with this process is illustrated in Fig. \ref{fig:TAC}. We also checked and made sure that there were no issues with the subtracted telluric sodium from the sky spectra retrieved with fiber B.

% Telluric correction figure

% Masters by night
\begin{figure*}[t]
\resizebox{\hsize}{!}{\includegraphics{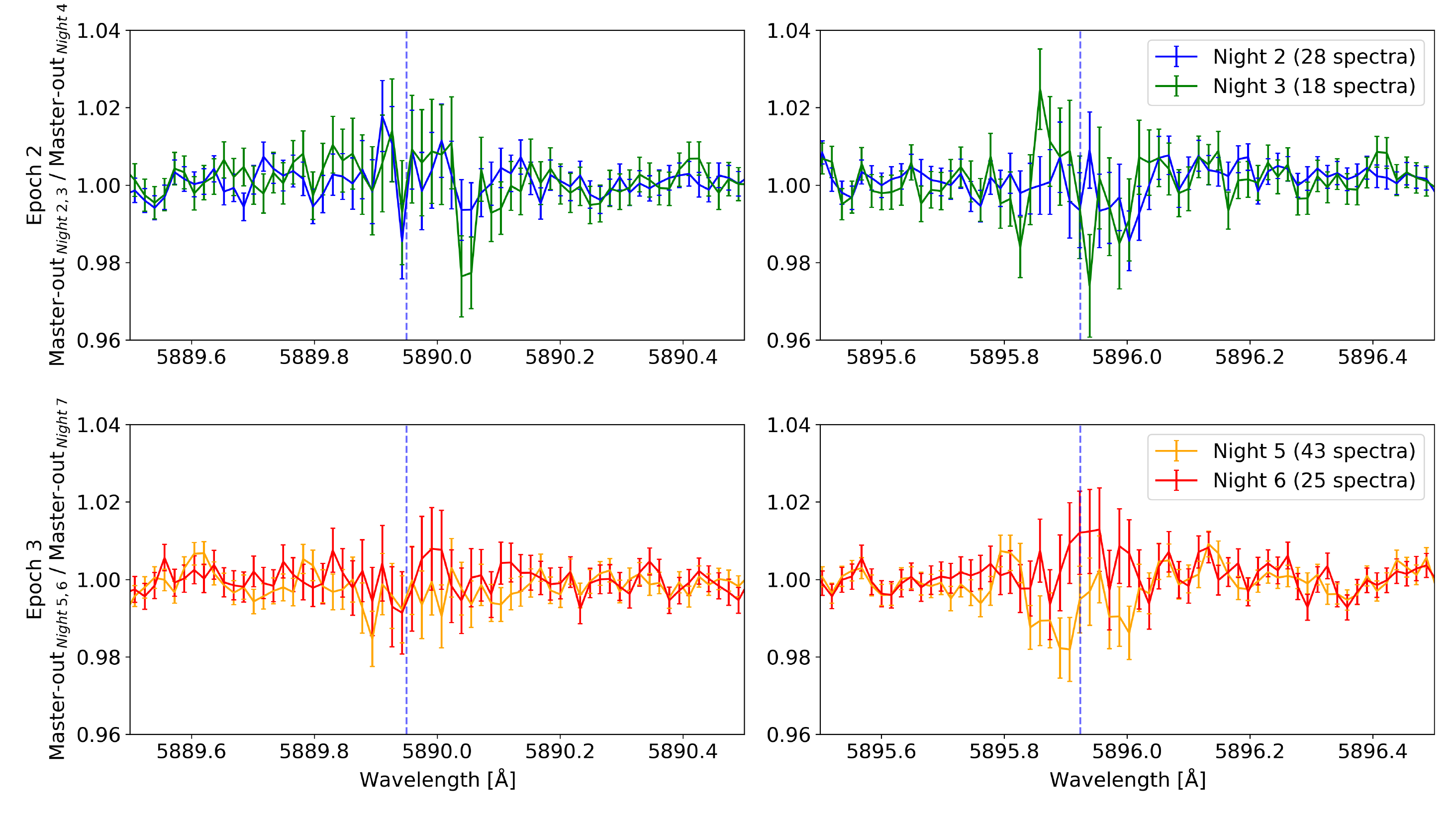}}
\caption{Masters-out of each night divided by the highest signal-to-noise master-out from their epoch (Night 4 and 7 respectively), zoomed on each line core of the Na D doublet. This shows the discrepancy between each night master within their respective epoch compared to the noise level. The center of each line is represented by a dashed vertical blue line. \textbf{Top}: Night masters residuals of Epoch 2. \textbf{Bottom}: Night masters residuals of Epoch 3. \textbf{Left}: Na D2 line zoom. \textbf{Right}: Na D1 line zoom.}
\label{fig:night}
\end{figure*}

% TRANSMISSION SPECTROSCOPY
\section{Transmission spectroscopy}

In this section, we describe how we derive the transmission spectrum \citep{SS2000} from the HARPS dataset following the methodology described in \citet{Wytt2015, Wytt2017}, adapted from \citet{Red2008}. We refine the procedure by adding transit depth and limb-darkening corrections for an accurate mapping of excess absorption into planetary radius variations. The successive processing steps are described below.

\paragraph{\textbf{Master spectrum computation}} First, the HARPS spectra are corrected for telluric contamination and shifted to the KELT-11 stellar rest frame by taking into account the barycentric correction, the systemic velocity and the star's radial velocity (RV) induced by KELT-11\,b. We determine if the recorded spectrum is in- or out-of-transit by comparing the time from the middle of the exposure to the orbital parameters given in Table \hyperlink{table.1}{1}. 

\paragraph{\textbf{Spectrum normalization}} We normalize the master spectrum and all in-transit spectra by computing a running average at every wavelength point and dividing by it. We choose a running mean window of 120 \AA, a large enough interval to avoid biasing of the continuum level by strong stellar absorption lines. The goal of the normalization process is to remove the uncalibrated broadband flux variations induced by Earth's atmosphere and instrumental flux losses. Normalized spectra are denoted by $\tilde{F}$ in the remainder of this paper.

\paragraph{\textbf{Transmission spectrum}} To obtain the transmission spectrum, the usual method is to compute the quantity $\tilde{F}_\mathrm{in,i} /  \tilde{F}_\mathrm{out} - 1$, which is meant to represent the wavelength-dependent absorption by the planet atmosphere that is in excess of the broadband absorption by the planet opaque disk. We refer to this widely-used quantity as "excess absorption". Here we revisit this formula and derive a more accurate expression for the wavelength-dependent planetary radius $R_\mathrm{p}(\lambda)$, in a similar fashion to \citealt{Wytt2020}. We start by explicitly writing the relation between out-of-transit and in-transit spectral fluxes, using the normalized spectra:

\begin{equation}
(1-\delta_\mathrm{i}) \tilde{F}_\mathrm{in,i}(\lambda) = \tilde{F}_\mathrm{out}(\lambda) - \frac{R_\mathrm{p}^2(\lambda)}{R_\mathrm{*}^2} \frac{LD_\mathrm{i}}{LD_\mathrm{mean}} \tilde{F}_\mathrm{local,i}(\lambda)
\label{eq:1}
.\end{equation}

In this equation, $\delta_\mathrm{i}$ is the fractional, broadband absorption caused by the transiting planet at orbital phase $i$, that is the quantity $1-\delta_\mathrm{i}$ is the broadband ("white-light") transit light curve, which is assumed to be known from precise photometric observations. $LD_\mathrm{i}$ and $LD_\mathrm{mean}$ represent the stellar limb darkening at the position of the transiting planet and the disk-averaged limb darkening, respectively. Finally, $\tilde{F}_\mathrm{local,i}(\lambda)$ is the normalized local stellar spectrum occulted by the planet at phase $i$. Equation~\ref{eq:1} expresses the fact that the light blocked by the planet at any wavelength is equal to the fractional planet disk size at that wavelength multiplied by the local stellar spectrum. We note that Eq.~\ref{eq:1} only applies to fully in-transit orbital phases (ingress and egress are not considered here).

Equation~\ref{eq:1} can be inverted to isolate the planetary radius:

\begin{equation}
\frac{R_\mathrm{p}^2(\lambda)}{R_\mathrm{*}^2} = \frac{LD_\mathrm{mean}}{LD_\mathrm{i}} \frac{\tilde{F}_\mathrm{out}(\lambda) - (1-\delta_\mathrm{i}) \tilde{F}_\mathrm{in,i}(\lambda)}{\tilde{F}_\mathrm{local,i}(\lambda)}
\label{eq:2}
.\end{equation}

At this point, a simplification can only be made if we assume that the occulted local stellar spectrum $\tilde{F}_\mathrm{local,i}(\lambda)$ is equal to the disk-integrated stellar spectrum $\tilde{F}_\mathrm{out}(\lambda)$. This amounts to neglecting the RM effect and CLVs in stellar line profiles. Such a simplification cannot be made in general, and RM+CLV effects must be taken into account using various correction methods. For the specific case of the HARPS KELT-11 observations presented here, we show in Sect.~\ref{section:RM} and Fig. \ref{fig:RM_sim} that these effects can in fact be neglected and Eq.~\ref{eq:2} simplifies to:

\begin{equation}
\frac{R_\mathrm{p}^2(\lambda)}{R_\mathrm{*}^2} = \frac{LD_\mathrm{mean}}{LD_\mathrm{i}} \left(1 - \frac{(1-\delta_\mathrm{i}) \tilde{F}_\mathrm{in,i}(\lambda)}{\tilde{F}_\mathrm{out}(\lambda)} \right)
\label{eq:3}
.\end{equation}

We note that this formula is close, but not equal to, the usual expression for the excess absorption $\tilde{F}_\mathrm{in} /  \tilde{F}_\mathrm{out} - 1$ (with inverted sign as radius is expressed instead of absorption). The main difference comes from the recognition that limb darkening induces a smaller atmospheric absorption signal at phases close to the stellar limb compared to the stellar disk center, and this must be corrected for to retrieve the phase-independent planetary radius.

We apply Eq.~\ref{eq:3} to all KELT-11 in-transit spectra to obtain the individual transmission spectra. We compute KELT-11\,b's light curve $1-\delta$ and $LD$ values using the Python package batman \citep{Kreid2015} and system parameters given in Table~\ref{table:params}.

\paragraph{\textbf{Shift to the planetary rest frame}} The main modification to \citet{Red2008} from \citet{Wytt2015} is to account for the change of the radial velocity of the transiting planet, which varies from -27 km s$^{-1}$ to +27 km s$^{-1}$ during the transit. This velocity change corresponds to a $\sim$ 1 \AA\, shift from blue to red for the planetary sodium lines we are probing. This effect is removed by shifting the individual transmission spectra to the planetary rest frame.

\paragraph{\textbf{Merging and weighting}} We merge the individual transmission spectra by averaging them for each epoch using a weighted mean. We do the same when merging epochs together to get the final transmission spectrum. Weighting allows us to properly take into account the S/N across wavelength regions and its temporal evolution. The weights are computed from two components: a spectral part and an exposure-related part. The spectral weight comes from the master-out for each epoch. Its goal is to account for the strongly variable S/N across continuum regions and deep lines in the stellar spectrum. The exposure-related weight comes from the integrated flux in each spectrum, in order to account for global flux variations across the nights. This process is described in more details in Bourrier et al. (in prep.).

\paragraph{\textbf{Sodium lines coaddition}} Since we observed full transits, we can trace the evolution of the absorption in the sodium doublet through time as the planet crosses in front of its host star. To illustrate this, we plot a map of the excess absorption as a function of orbital phase and wavelength in the range of the sodium lines (see \citealt{Allart2018}), where we bin together spectra with similar phases. To further enhance our perception of the absorption lines, we coadd both lines of the sodium doublet in order to increase the S/N. To do so, we transform the wavelength scale around the sodium lines to a velocity scale centered on the laboratory wavelength of each line. Then we average the absorption values of the pixels with the same velocity. We scan through a radial velocity range of -150 to 150 km s$^{-1}$ in order to follow the Doppler-shifted sodium lines through 3 days of consecutive observations with a 0.82 km s$^{-1}$ step, which is the average pixel size on the HARPS CCD. At 5890 \AA, this corresponds to a step of 0.016 \AA. The wavelengths chosen as the zero velocity are the two theoretical rest frame wavelengths of the Na doublet (Na D1 at 5895.924 \AA, Na D2 at 5889.950 \AA).
 
\begin{figure*}[h]
\resizebox{\hsize}{!}{\includegraphics{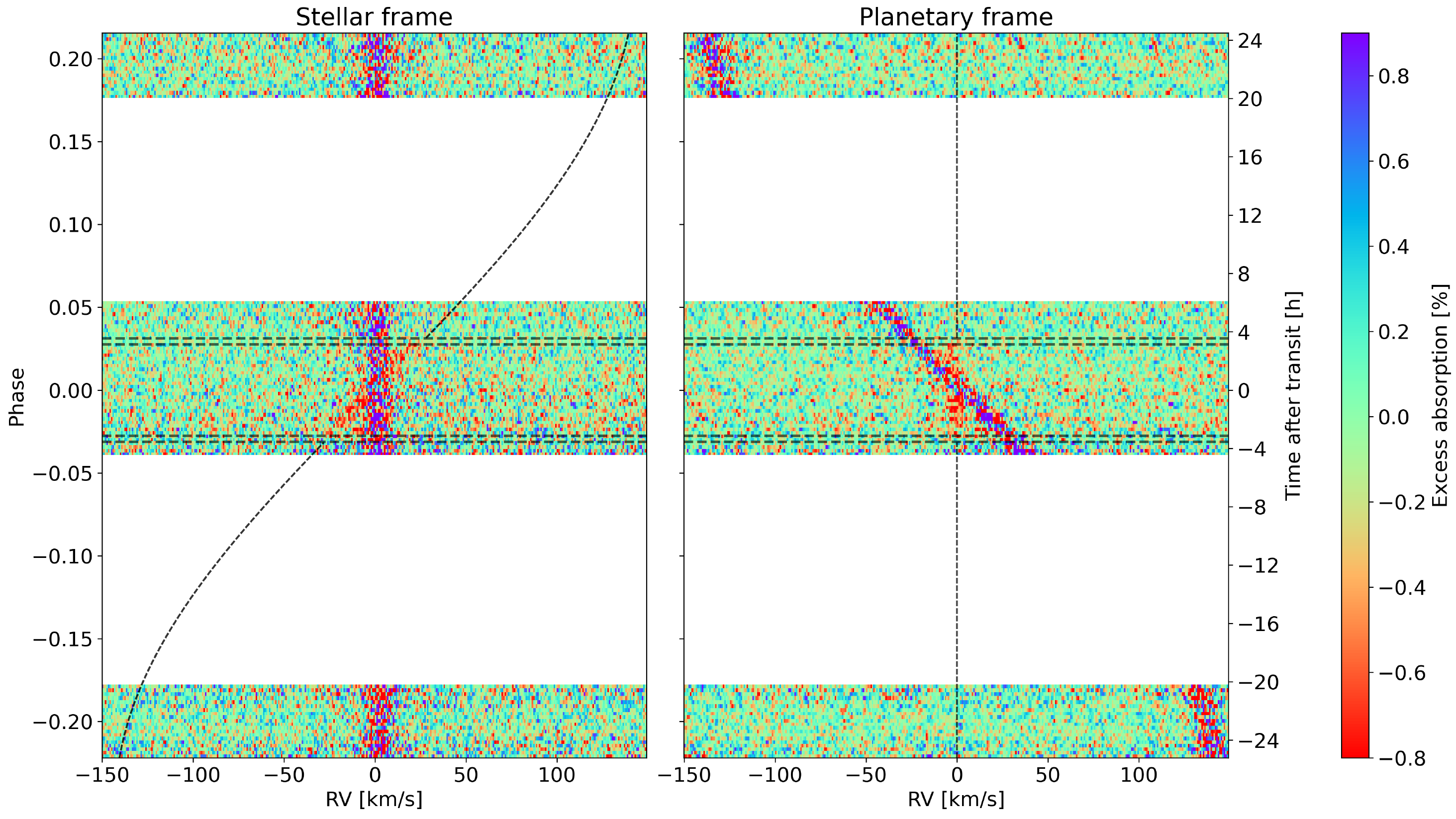}}
\caption{Excess absorption maps of KELT-11\,b as a function of orbital phase, binned every 0.2\% in phase, with both lines from the Na D doublet coadded. The dashed horizontal black lines indicate the transit contact points $T_\mathrm{1}$ to $T_\mathrm{4}$ from the start (bottom) until the end (top) of the transit. The dashed horizontal/diagonal black lines represent where the stacked sodium line should appear according to the orbital velocity of the planet; they are not traced inside the transit. \textbf{Left}: Excess absorption maps in the stellar rest frame. \textbf{Right}: Excess absorption maps in the planetary rest frame.  Excess absorption in the sodium lines originating in the planetary atmosphere is clearly visible as a red signature over the duration of the transit. The thick multicolor traces correspond to the noisy parts of the stellar spectrum in the core of the Na D lines.}
\label{fig:stack}
\end{figure*}

This process is a simplification of the cross-correlation function that is generally used for coadding tens or more lines together. The result is a two-dimensional map of the absorption as a function of orbital phase and radial velocity, either in the stellar rest frame or in the planetary rest frame if the spectra have been shifted accordingly (Fig. \ref{fig:stack}).

% RESULTS
\section{Sodium transmission spectrum of KELT-11\,b}
\label{section:4}

% Individual Masters / Total Master
\subsection{Masters comparison}
We first build the master-out from each night. The masters need to be coherent with one another in the same epoch in order to justify combining them. To be able to see the differences between masters, we plot each night master divided by the highest signal-to-noise master from that epoch and show their variations around the sodium doublet in Fig. \ref{fig:night}. The masters in each epoch seem generally compatible within their error bars. The only visible discrepancy concerns the sodium D1 line in Epoch 3, which may partly explain why this absorption line is less well defined than the D2 line in the individual epochs (Fig. \ref{fig:tsepoch}) and final transmission spectrum (Fig. \ref{fig:finalts}), but the variations are mostly below 1$\sigma$ and at most 2$\sigma$ around the Na D1 line, which is marginal.

We then merge all out-of-transit spectra of each observation epoch. Their S/Ns range from 118, 171 and 200 respectively in the sodium lines core to 495, 720 and 837 in the continuum. If we wanted to use only the spectra from the transit night, the S/N would drastically drop. For example, Epoch 3 would have a S/N of $\sim$ 70 in the line core and $\sim$ 295 in the continuum.

Combining out-of-transit spectra from all different epochs could also be a possibility, but the several weeks between each epoch may result in strong stellar variability and differences in observational conditions. To test this possibility, we computed the ratio between the masters-out of Epoch 2 and Epoch 3. We note that variations in this ratio around the line cores are somewhat larger than within epochs, which we interpret as evidence for stellar variability over weekly time scales. Thus, we decided not to combine master spectra across epochs.

% 2D Absorption Map
\subsection{2D absorption map}
Before combining in-transit spectra, we compute and show all individual spectra as a function of the orbital phase. We coadd both lines from the sodium doublet. This boosts the S/N, which is needed to resolve the excess absorption since the sodium lines core low S/N dominates the binned transmission spectra. The spectra that have roughly the same orbital phase are merged together within bins of 0.2 \% in phase, which correspond to around 14 minutes.  This results in a two-dimensional map (see Fig. \ref{fig:stack}) where the absorption is color-coded with the wavelength and the orbital phase as the X and Y axis respectively. A less than zero value means excess absorption in percent. 

We plot this map in the stellar and the planetary rest frames. In the stellar rest frame, we can see the increased noise in the center of the stellar sodium lines. The planetary sodium absorption signature clearly appears as a red trace when the planet starts transiting and stops when it exits transit. 

% Transmission spectrum
\subsection{Transmission spectrum}

\begin{figure*}
\resizebox{\hsize}{!}{\includegraphics{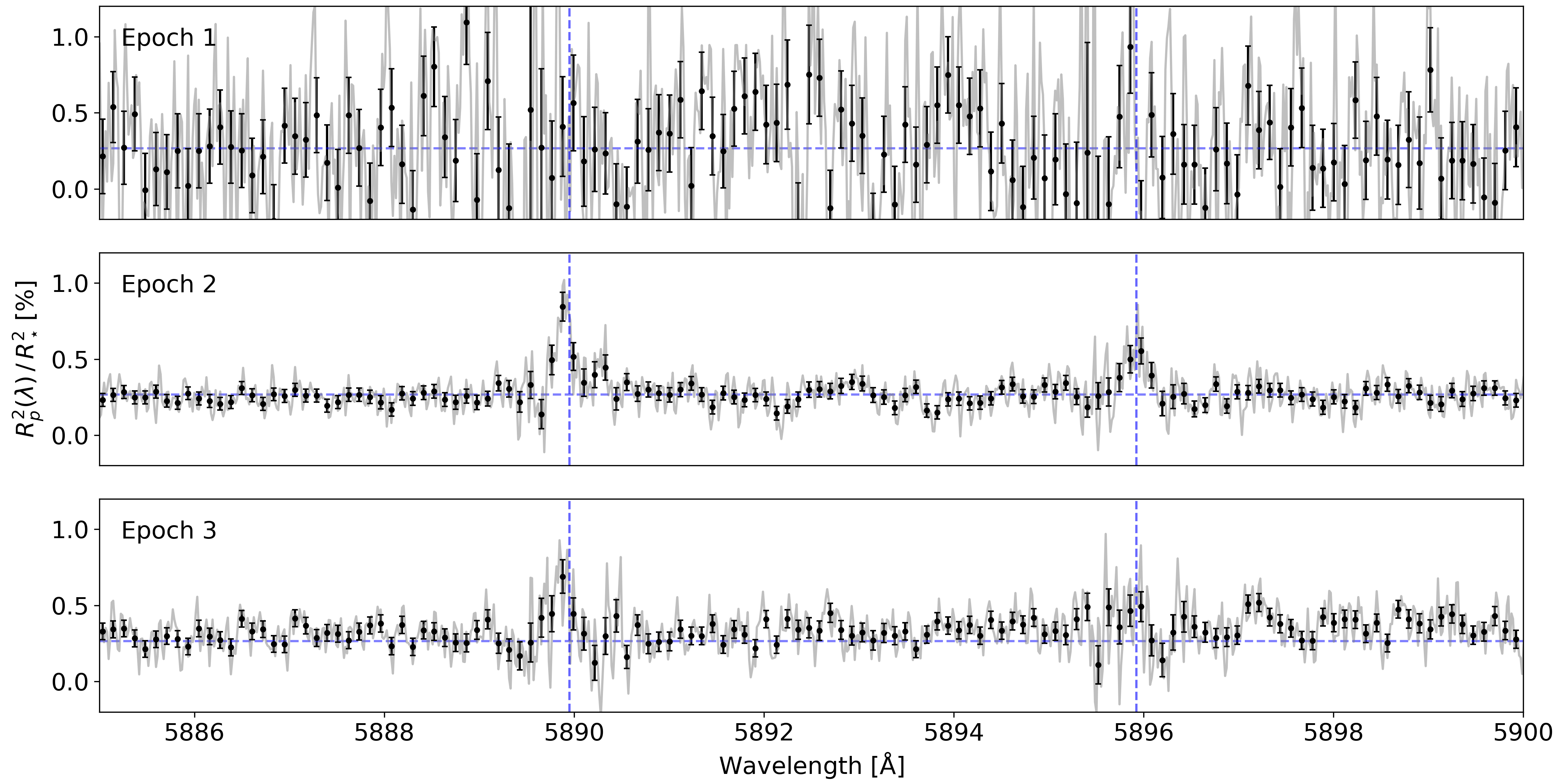}}
\caption{Transmission spectrum of KELT-11\,b in the Na doublet wavelength range for each transit observed with HARPS, at the same scale. The vertical blue dashed lines show the planetary rest frame transition wavelength of the Na doublet, the horizontal one is the white-light radius $R_p^2 / R_*^2$. The gray line represents the unbinned resulting transmission spectrum. The black points show the binned spectrum with a 0.1 \AA \,step.}
\label{fig:tsepoch}
\end{figure*}

We computed a transmission spectrum $\frac{R_p^2(\lambda)}{R_*^2}$ of KELT-11\,b for each epoch of observations using Eq. \ref{eq:3} which can be seen in Fig. \ref{fig:tsepoch}. We see an excess absorption in Epochs 2 and 3 for the sodium doublet. Epoch 1 is much more noisy, because of the low number of in-transit spectra (3). Therefore, we choose not to use the Epoch 1 spectra when combining each spectrum of the different epochs. 

\begin{figure*}[h!]
\resizebox{\hsize}{!}{\includegraphics{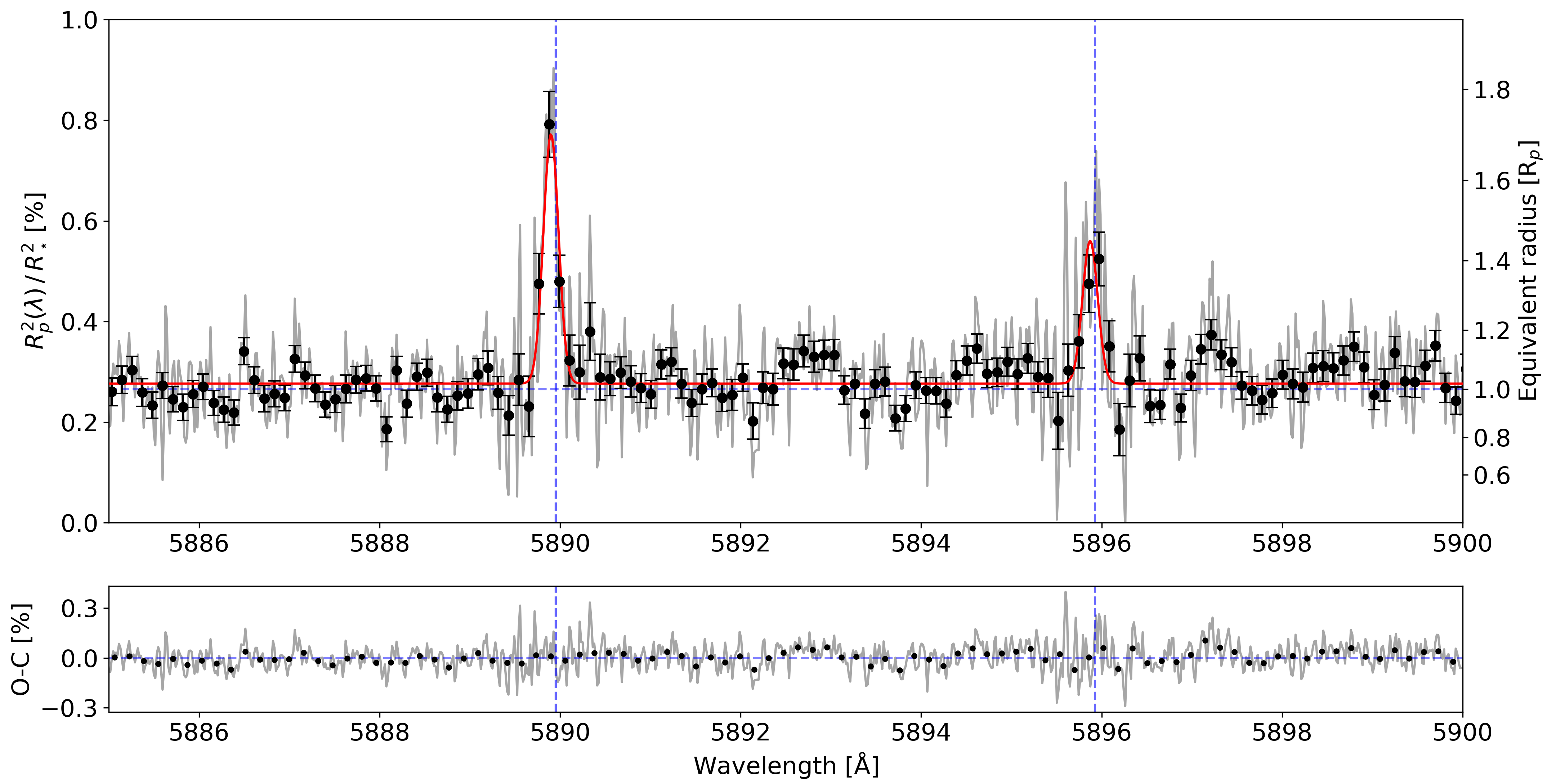}}
\caption{Combined transmission spectrum of KELT-11\,b in the sodium doublet wavelength range, using Epoch 2 and Epoch 3 merged from HARPS observations.  The vertical blue dashed lines show the planetary rest frame transition wavelength of the Na I doublet, the horizontal one is the white-light radius. \textbf{Top}: Transmission spectrum and Gaussian fit of the lines. The gray line represents the unbinned resulting transmission spectrum. The black points show the binned spectrum with a 0.1 \AA \,step. The red line is the Gaussian fit to the unbinned transmission spectrum. \textbf{Bottom}: Residuals from the Gaussian fit to the transmission spectrum. The gray line represents the residuals at each wavelength and is binned in the same way as the transmission spectrum.}
\label{fig:finalts}
\end{figure*}

The spectrum from Epoch 2 (Fig. \ref{fig:tsepoch}) shows a clear absorption trace in the core of the sodium lines. The transmission spectrum obtained during Epoch 3, also shows clear absorption in the D2 line, but is noisier in the D1 line. In order to boost the S/N, we merge the Epochs 2 and 3 transmission spectra,  averaging them with their respective weights. The resulting spectrum is shown in Fig. \ref{fig:finalts}. We perform a Gaussian fit on the absorption lines of the sodium doublet by fixing the distance between both lines. The free parameters are the amplitude of each line, the center of the lines (with a fixed distance between them), the full width at half-maximum (FWHM) which is assumed to be the same for each absorption line and an offset compared to the continuum (which absorbs any residual normalization  error). The results of the fit are displayed in Table \ref{table:gauss}.

% DISCUSSION
\section{Discussion}
\subsection{Sodium absorption detection}
The sodium absorption signature is clearly detected in the final transmission spectrum with a low full width at half maximum of 0.03 $\AA$. The depth of the absorption is 0.50 $\pm$ 0.06 \% for the D2 line and 0.28 $\pm$ 0.05 \% for the D1 line. This corresponds to a 8$\sigma$ and 6$\sigma$ detection respectively. These absorptions correspond to extended radii of 1.69 and 1.43 times the planetary white-light radius. These absorptions probe around 24 and 15 scale heights respectively for the D2 and D1 lines. 
We simulated the impact of the rotation of the star to ensure that the sodium signature does not come from the Rossiter-Mclaughlin effect. However, KELT-11 is as slow rotator with a projected rotation speed of 1.99 km s$^{-1}$, so the Rossiter-McLaughlin contribution is completely lost in the noise of the spectrum as shown in Appendix \ref{appendix:RM}. The Rossiter-McLaughlin effect is studied in detail using the reloaded Rossiter-McLaughlin model \citep{Cegla2016, Bourrier2018} in Sect. \hyperlink{section.6}{6} and we simulate the magnitude of this effect on the merged transmission spectrum in Fig. \ref{fig:RM_sim}.

The difference between our calculated line center and the theoretical one is a blueshift of -0.05 $\pm$ 0.01 \AA. This difference corresponds to a radial velocity of -2.5 $\pm$ 0.5 km s$^{-1}$, which may be explained by winds from the day side to the night side. We explore this possibility by modeling different wind patterns using MERC \citep{Seidel2020a, Seidel2021} in Sect. \hyperlink{section.7}{7}.

% THREE DAYS OBS DISCUSSION
\subsection{Robustness of 3-day observations}
As of today, the main tool for chemical detections in the atmosphere of exoplanets is still the transmission spectrum, which yields its best results with high-resolution spectroscopy. The use of ground-based telescopes is primordial to further advance atmospheric characterization. But these telescopes have limitations compared to their space-based counterparts: they are dependent on the day-night cycle, which means telescopes cannot operate for most of the duration of the day because of sunlight, and the Earth's atmosphere blurs the incoming signals. While the latter can be solved by precise telluric correction and adaptive optics, the former will always be an issue. Observing runs cannot last more than the duration of a night, which represents around 8 hours per day at best. As many out-of-transit spectra from before and after a transit are needed to build a robust baseline, the transit duration is limited to only a few hours. This means that chemical signatures from the ground can mostly be detected on short-transiting exoplanets. Up to now, the planets with the longest transit durations where sodium was detected stood at around 4.5 hours (WASP-17b, \citealt{Sing2016} from space and WASP-127b, \citealt{Chen2018, Allart2020}, from the ground), compared to the 7 hours long transit duration of KELT-11\,b.

The 3-day observation method allows us to bypass this transit duration issue. With it, we can build precise masters-out spectra during the day-before and day-after observations, as recording many spectra helps boosting the S/N. The second night is used as the in-transit spectra collector, and planets with longer transit duration can gather more in-transit spectra to analyze, again boosting the S/N of the transmission spectrum. As we saw in Sect. \ref{subsection:dace}, the night-by-night variations are mostly negligible when combining the out-of-transit spectra from each night to construct the master-out, and while the transmission spectra for each epoch contain some level of red noise, the quality of the master-out and transmission spectra show that this method is stable and robust.

It is remarkable that the 7-hour transit of KELT-11\,b was observed from start to finish in an 8 hours observation session twice. For future observations, knowing the transit epoch and duration of exoplanetary transits with precision will be primordial in order to correctly flag in- and out-of-transit spectra. 

Of course, there are limitations to the use of this method. The main obstacle would be an intrinsically active star. It would render the use of this technique very difficult. The observing conditions must stay stable for the whole length of the data recording, as one night of bad weather -- especially on transit night -- could render the whole set of observations unusable. Observations may also suffer from technical problems that would affect the observations in a similar way as bad meteorological conditions.  Finally, more effort must be put in verifying that all spectra and masters show no significant discrepancy between them. As suggested by Fig. \ref{fig:night} and Fig. \ref{fig:tsepoch}, discrepancies in the day-by-day masters could affect the final transmission spectrum. All in all, it is a higher risk with higher reward strategy, that can provide information on targets not observed from the ground until now.

\begin{table}[t]
\centering
\caption{Results of Gaussian fit on combined transmission spectrum from Fig. \ref{fig:finalts}.}
\label{table:gauss}
\begin{tabular}{l l}
\toprule
Parameter & Fitted value\\
\midrule
Na D2 line depth & 0.50 $\pm$ 0.06 \% \\
Na D1 line depth & 0.28 $\pm$ 0.05 \% \\
Line center shift & -0.05 $\pm$ 0.01 $\AA$ (-2.5 $\pm$ 0.5 km s$^{-1}$) \\
FWHM & 0.14 $\pm$ 0.02 $\AA$ (7.1 $\pm$ 0.8 km s$^{-1}$) \\
Continuum offset & 0.011 $\pm$ 0.003 \%\\
\bottomrule
\end{tabular}
\end{table}

\subsection{Comparison with previous KELT-11\,b results} 
While no sodium signature was detected in their KELT-11\,b transmission spectrum, \citet{Zak2019} used the same HARPS data set and followed \citet{Wytt2015} in order to compute the transmission spectrum. The telluric correction method used follows \citet{CB2017}, which consists of comparing the observed spectra with a telluric water model to subtract the telluric contamination. We did not find any significant difference in the resulting telluric correction between this method and our use of \textsc{Molecfit}. The normalization method chosen is also different, but should not be responsible for the excess absorption. The fact that all spectra were used to build one and only master-out might be a reason, as discrepancies in between epochs could lead to a dilution of the signature, but it seems unlikely that the variations between spectra of different epochs would cause such a difference. 

We believe the main cause of the divergence in results to be the choice of transit ephemeris: in \citet{Zak2019}, the main source of system parameters was \citet{Pepp2017}. These were subsequently refined by \citet{Beatty2017} and \citet{Colon2020} using new sets of observations. As explained in Sect. \ref{subsection:dace}, there is a significant offset in the transit times which changes greatly which spectra are considered in- and out-of transit, which is more likely to explain the nondetection in \citet{Zak2019}.

The work of \citet{Colon2020} indicates the presence of water vapor in KELT-11\,b's atmosphere, albeit in abundances lower than expected for this kind of planet, as well as HCN, TiO and AlO depending on the model chosen. \citet{Changeat2020} confirms the low abundance water vapor signature and adds the consistent detection of CO$_2$ with possibly other carbon-bearing molecules (CO, HCN). The SPITZER secondary eclipse analysed by \citet{Colon2020} suggests a redistribution of heat from the day-side to the night-side, which corroborates our wind modeling scenario. In any case, the main challenges of KELT-11\,b's characterization will be the determination of its clouds and haze properties and further observations from high-end facilities are needed to constrain those parameters and the choice of models, paramount to determining the planet composition.

\begin{figure}[h]
\resizebox{\hsize}{!}{\includegraphics{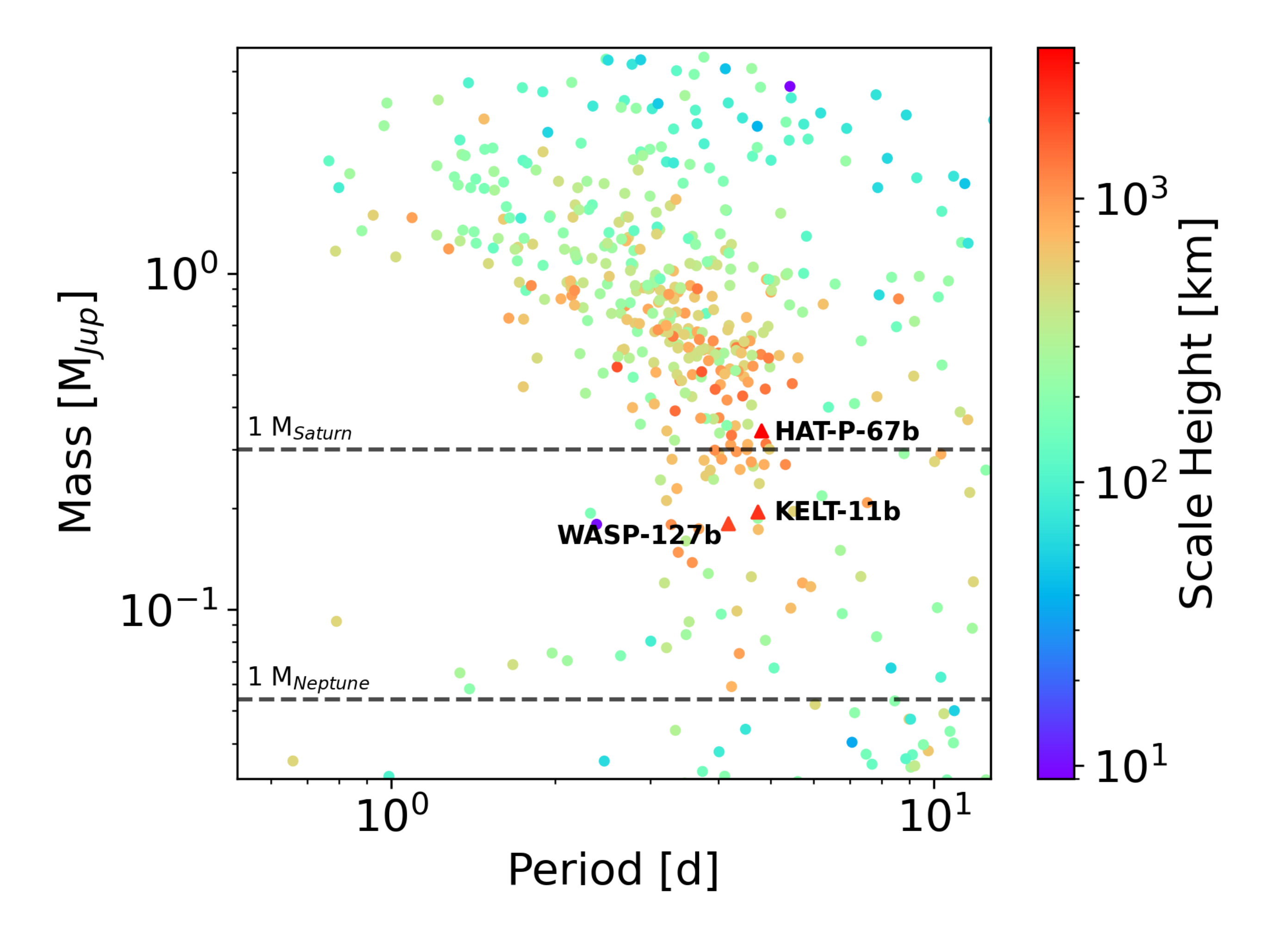}}
\caption{Period -- mass diagram for known exoplanets around the hot Neptune desert color-coded with their calculated scale height value, assuming a H$_2$+He atmosphere. Highlighted are the three planets in this region with a scale height of more than 2000 km.}
\label{fig:P-M diag}
\end{figure}

\subsection{Comparison to similar planets}
KELT-11\,b belongs to the peculiar group of inflated sub-Saturns with a scale height of H = 2763 km, an equilibrium temperature of $T_\mathrm{eq}$ = 1712$^{+51}_{-46}$ K with a bright metal-rich subgiant host with metallicity of [Fe/H] = 0.17 $\pm$ 0.07 \citep{Beatty2017, Pepp2017}. This alone is a great example of the diversity of exoplanets and make it a great target for atmospheric characterization. Since inflated sub-Saturns are not represented in our Solar System and a recently discussed group of planets, they are important in order to understand and compute models for planet formation. They challenge our understanding of the runaway gas accretion phase that forms the giant gaseous planets.

A few sub-Saturns have already been studied with transmission spectroscopy. In Fig. \ref{fig:P-M diag}, we show the different magnitudes of scale heights of exoplanets around the hot Neptune desert. Their scale height is calculated using $H = \frac{RT}{\mu g}$, assuming a hydrogen and helium atmosphere ($\mu = 2.4$ g mol$^{-1}$) and $g = \frac{G M_\mathrm{P}}{R_\mathrm{P}^2}$. Very few sodium signatures have been detected in and around the Neptune desert, most recently on WASP-166\,b \citep{Seidel2022}.

For WASP-127b, one of the only exoplanets with a scale height of the same order of magnitude as KELT-11\,b (see Fig. \ref{fig:P-M diag}), a curiously large sodium feature was originally found in \citet{Zak2019} in HARPS data which was later attributed to stellar contamination in \citet{Seidel2020c}. The upper limit on a possible sodium signal provided in \citet{Seidel2020c} for the HARPS HEARTS dataset is consistent with the subsequent sodium detection with ESPRESSO presented in \citet{Allart2020}. While it is supposed to have a haze layer, absorption traces of potassium, lithium, carbon dioxide and water vapor were also found on WASP-127\,b \citep{Lam2017, Chen2018, Welbanks2019, Zak2019}. 

WASP-39\,b \citep{Faedi2011, Mancini2018} orbits a metal-poor star and has a large scale height of 983 km with an equilibrium temperature of 1166 $\pm$ 14 K. While whether its atmospheric metallicity is still debated, sodium, potassium and water signatures have been detected in its atmosphere \citep{Wakeford2018, Kirk2019}. 

In similar fashion to WASP-127b, WASP-69\,b \citep{And2014} has similar mass, radius and period as KELT-11\,b. Sodium and helium absorptions were detected using observations from HARPS-N \citep{CB2017} and CARMENES \citep{Nort2018, Khal2021}. Comparatively to KELT-11\,b, it exhibits a scale height of $\sim$ 650 km and a lower equilibrium temperature of 963 K, with a much higher sodium absorption of 3.2 and 1.2 \% in the D2 and D1 lines respectively \citep{Khal2021}.

As mentioned earlier, \citet{Colon2020} and \citet{Changeat2020} showed water absorption on KELT-11\,b as well, with possible traces of oxygen- and carbon-bearing molecules. It seems that while determining their metallicity still is tricky, these planets share features such as water absorption lines, as well as some alkali metallic features such as sodium or potassium. Since the depth of sodium absorption detected on KELT-11\,b is lower than expected for such an inflated planet, it makes sense to assume that high-altitude clouds partially hide the supposedly feature-rich atmosphere of this kind of exoplanets.

% RELOADED RM
\section{Reloaded Rossiter-McLaughlin technique} \label{section:RM}

\begin{figure}
\resizebox{\hsize}{!}{\includegraphics{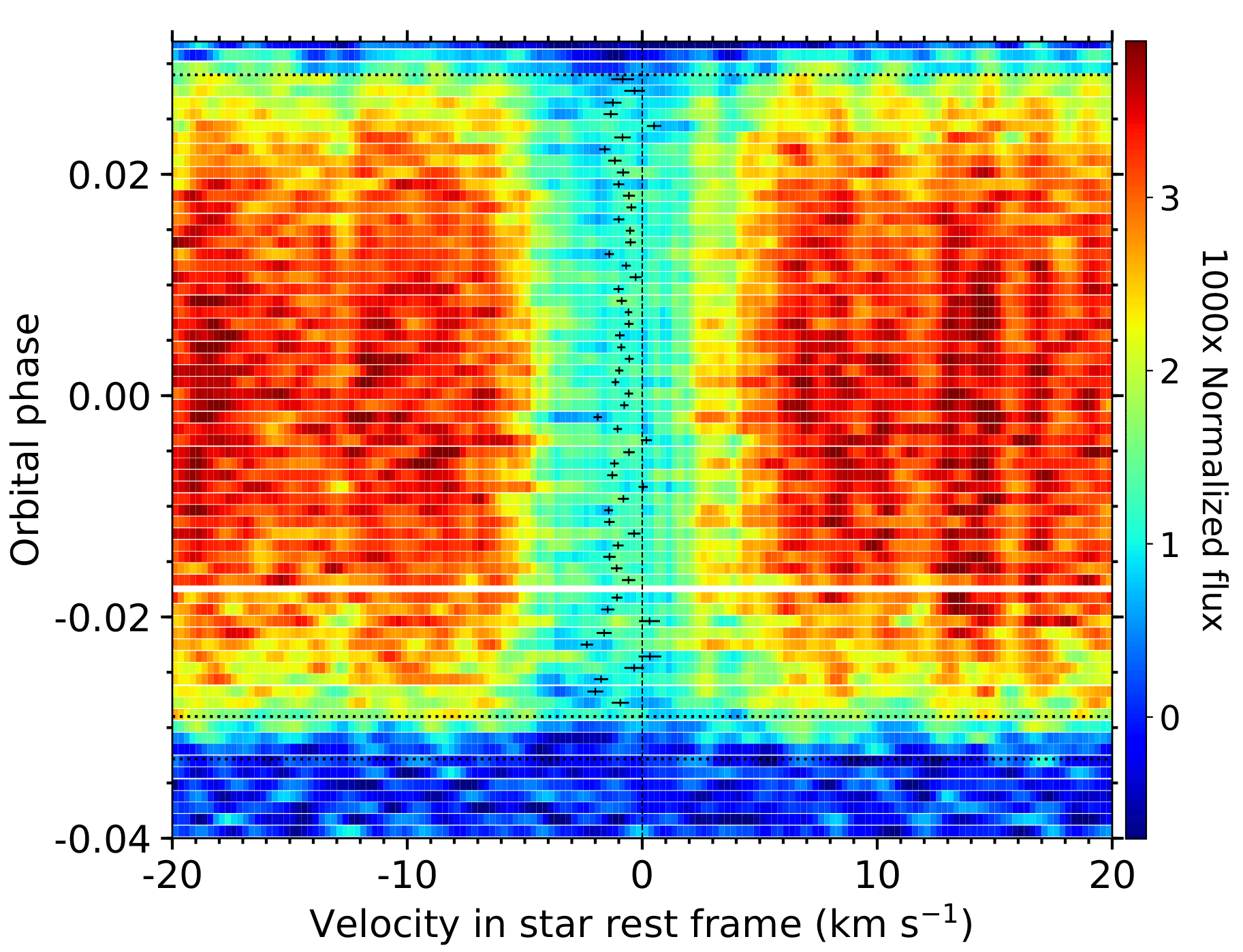}}
\resizebox{\hsize}{!}{\includegraphics{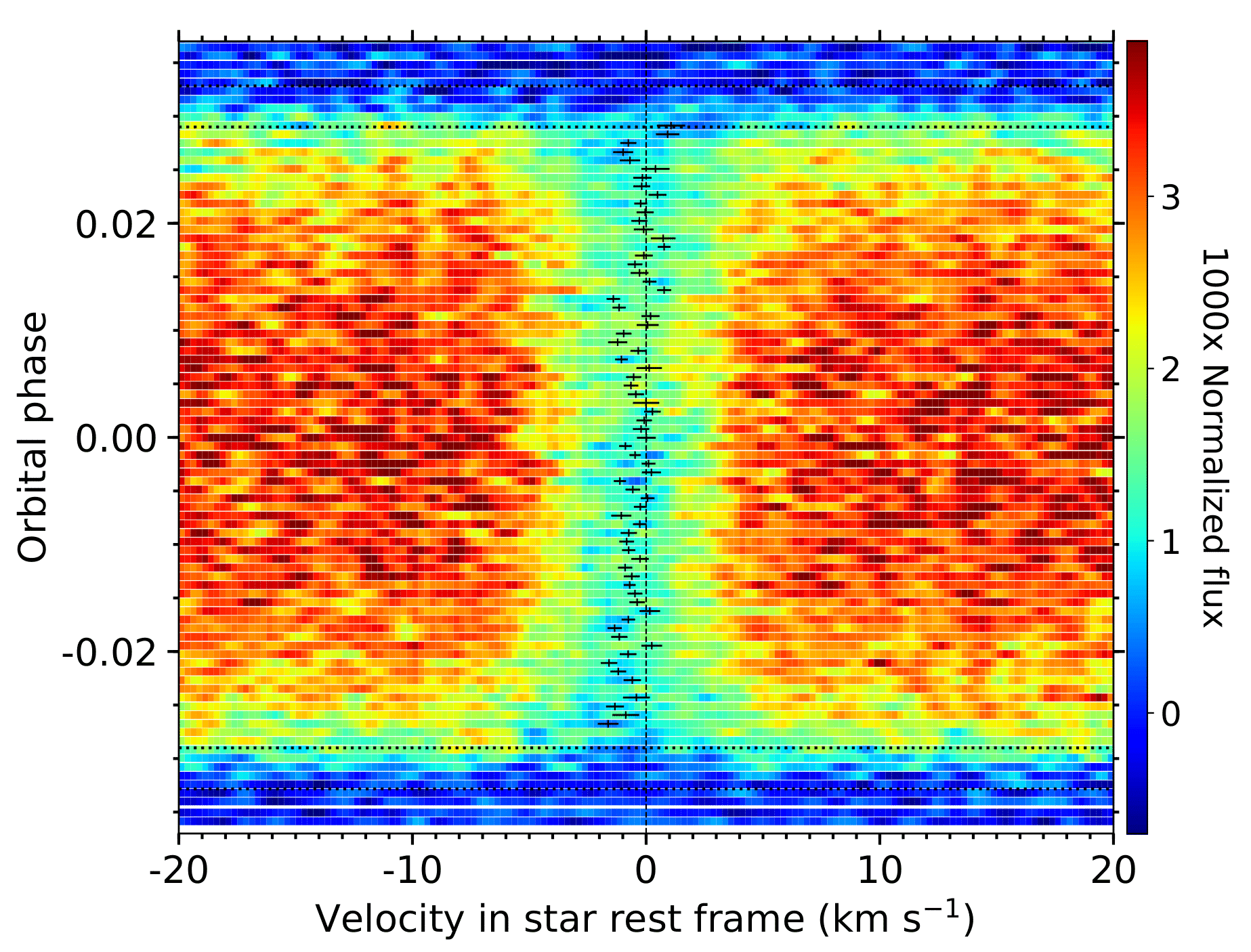}}
\caption{Map of the residual CCFs during the 16-02 (\textbf{top}) and the 07-03 (\textbf{bottom}) nights as a function of orbital phase and RV in the stellar rest frame. Colors indicate flux values. The horizontal dashed black lines indicate transit contacts. In-transit residual CCFs show the average stellar line profiles from the regions occulted by KELT-11\,b across the stellar disk. The crosses are the measured centroids of the CCFs, which correspond to the local RVs of the planet-occulted regions.}
\label{fig:2dmap}
\end{figure}

We applied the reloaded RM technique (\citealt{Cegla2016}; see also \citealt{Bourrier2018}) to the 2017 HARPS observations of KELT-11\,b. Two nights were used for the analysis: February 16 and March 07. CCFs computed with a G2 mask were first corrected for the Keplerian motion of the star induced by the planet. The CCFs outside of the transit were coadded to build a master-out CCF representative of the unocculted star for each night, whose continua were normalized to unity. The centroids of the master-out CCFs, derived with a Gaussian fit, were used to align the CCFs in the stellar rest frame. The continua of all CCFs were scaled afterwards to reflect the planetary disk absorption by KELT11-b, using a light curve generated by the \texttt{batman} package \citep{Kreid2015} and the parameters from Table \ref{table:params}. Residual CCFs were then obtained by subtracting the scaled CCFs from the master-out in each night (Figure \ref{fig:2dmap}).

No spurious features are observed in the residual CCFs out of the transit. Within the transit, the residual RM spectrally and spatially resolve the photosphere of the star along the transit chord. The residual CCFs are well fit with Gaussian profiles using a Levenberg-Marquardt least squares minimization, setting flux errors to the standard deviation in their continuum flux. The centroid of the fitted residual CCFs correspond to the local RVs of the planet-occulted regions. The detection of the average local stellar lines is set according to the criterion defined by \citet{Allart2017}, that is the amplitude of the model residual CCF must be at least three times larger than the dispersion in the measured residual CCF continuum. All the residual CCFs were detected except for most of the exposures at the ingress and egress of the transit.

The local RV series were fitted with the model described in \citet{Cegla2016,Bourrier2017} assuming solid-body rotation for the star. We sampled the posterior distributions of $v\sin{i_\star}$ and $\lambda$ using the Markov chain Monte Carlo (MCMC) package \texttt{emcee} \citep{FM2013}, assuming uniform priors. The best-fit values of these two parameters significantly differ from one night to another. We derive $v\sin{i_\star}=2.38^{+0.08}_{-0.09}$ km s$^{-1}$, $\lambda=-82.07^{+1.95}_{-1.76}{}^\circ$ for the 16-02 night and $v\sin{i_\star}=1.34^{+0.11}_{-0.10}$ km s$^{-1}$, $\lambda=-66.06^{+4.11}_{-4.23}{}^\circ$ for the 07-03 night, which corresponds to $\Delta{}v\sin{i_\star}=7.3\sigma$ and $\Delta\lambda=3.4\sigma$ between the two nights.

We note that for this RM effect analysis, we used the out-of-transit exposures of the night in question for each of the two visits. The inclusion of an out-of-transit baseline from nights other than the in-transit nights (see Sect. \ref{section:2}) results in even more significantly different $v\sin{i_\star}$ and $\lambda$ between the two nights. As shown in Sect. \ref{section:4}, the combination of in- and out-of-transit data from different nights provides good results for transmission spectroscopy, but has to be taken with caution for the analysis of the RM effect via high-precision radial velocities. In this case, it appears important to use out-of-transit CCFs that precisely reflect the state of the stellar surface during the transit. In order to do so, the best strategy for transits that are as long as KELT-11\,b's might be to observe the first half of the transit with sufficient pretransit baseline in a given night, and the second half with post-transit baseline in another night.

\begin{figure}
\resizebox{\hsize}{!}{\includegraphics{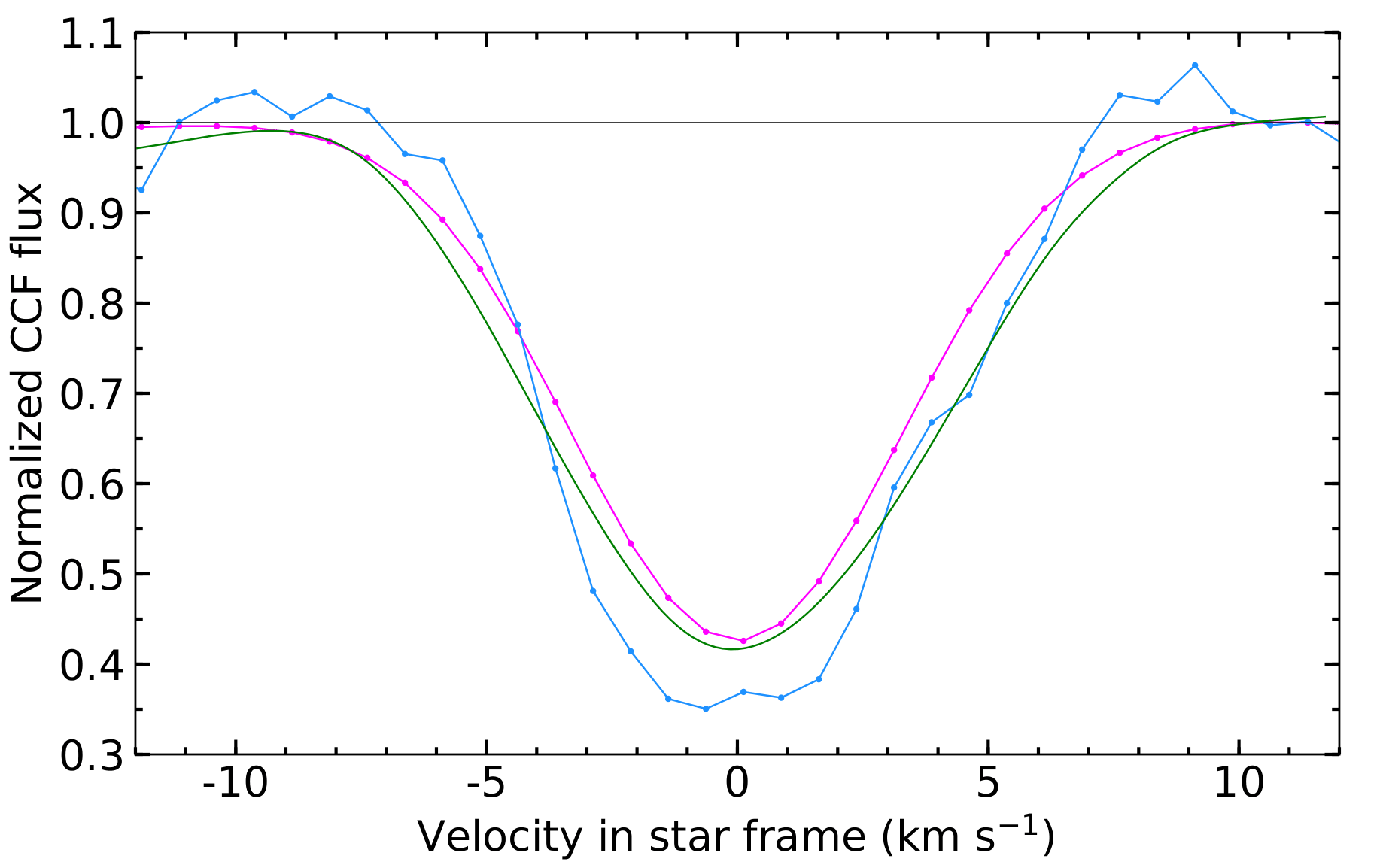}}
\resizebox{\hsize}{!}{\includegraphics{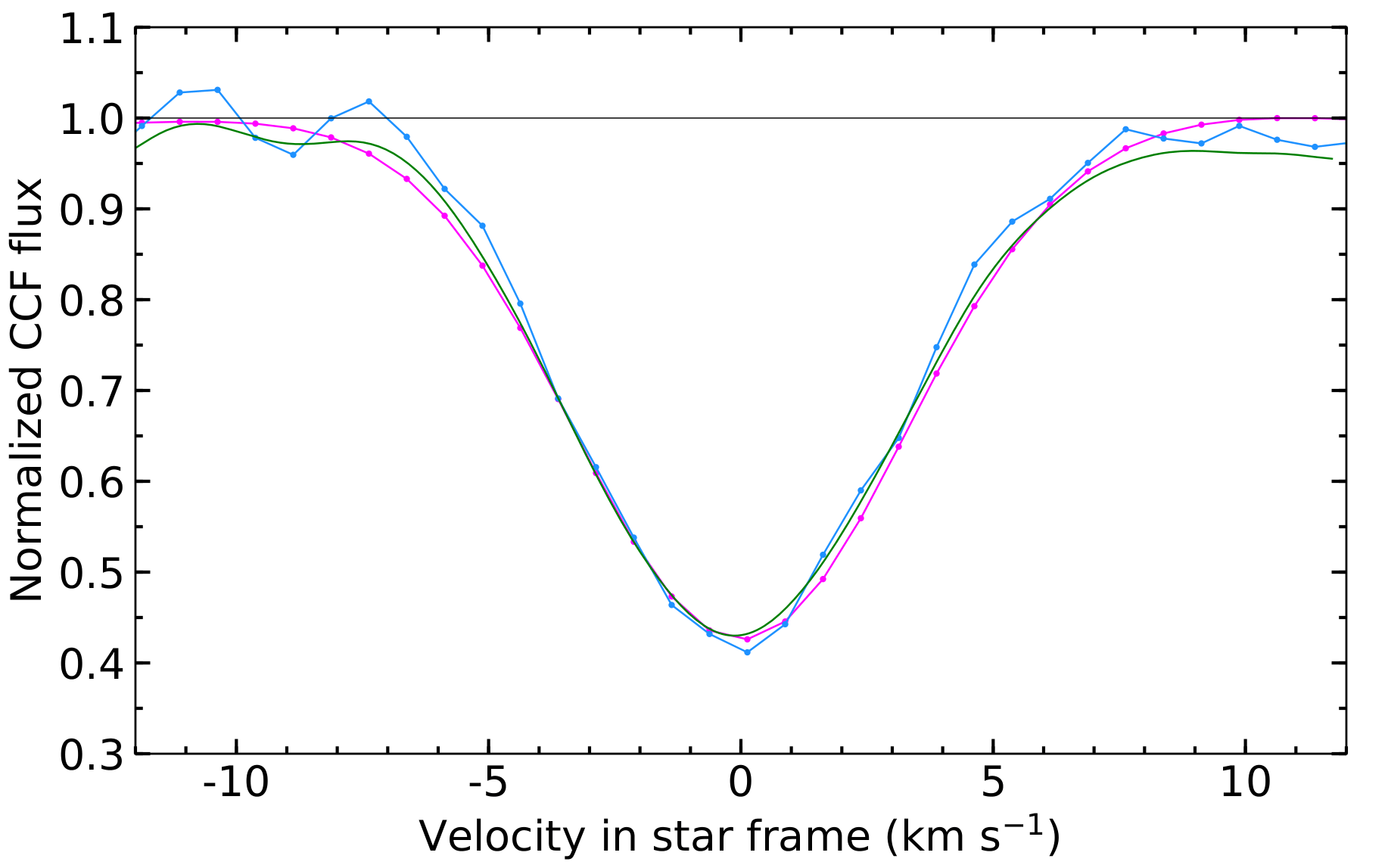}}
\caption{Master-out CCFs (fuschia lines) and their best fit (green lines) obtained by tiling a model star with the limb-darkened master-local CCFs used as proxies for the specific stellar intensity profile for the 16-02 (\textbf{top}) and the 07-03 (\textbf{bottom}) nights. The master-local CCFs are also showed in blue lines.}
\label{fig:mccfout}
\end{figure}

An estimate of $v\sin{i_\star}$ independent of the local RVs can be derived from the comparison between the master-out CCF and the master-local CCF \citep[see e.g.,][for a similar analysis of WASP-121 b]{Bourrier2020}. Under the assumption that the residual CCFs measured along the transit chord are representative of the entire stellar disk, the observed master-out was fit by tiling a model star with the limb-darkened master-local CCF, shifted in RV position by the solid rotation of the photosphere, which was let free to vary (Figure \ref{fig:mccfout}). The best fits correspond to $v\sin{i_\star}=4.16$ km s$^{-1}$ for the 16-02 night and to $v\sin{i_\star}=1.77$ km s$^{-1}$ for the 07-03 night. We obtain a good fit to the 07-03 night's master-out CCF, and the derived $v\sin{i_\star}$ is reasonably close to the one derived via the RM analysis. This suggests that the planet-occulted stellar profiles during this night are representative of the rest of the stellar surface and that the broadening of the local lines induced by the star rotation is the dominant mechanism in shaping the disk-integrated line. In contrast, we obtain a poorer fit to the 16-02 night's master-out CCF, and the derived $v\sin{i_\star}$ is substantially larger than the $v\sin{i_\star}$ derived for the other night. This is likely due to the deeper and narrower local CCFs for the 16-02 night, requiring a larger broadening to reproduce the master-out CCF (similar between the two nights). Furthermore, the derived $v\sin{i_\star}$ for the 16-02 night is also larger than the RM-derived value for this night. This suggests that, unlike the other night, the local stellar lines along the transit chord are not representative of the rest of the stellar surface. The planet-occulted local CCFs could be distorted enough that the measured centroids do not trace the stellar rotation, which results in a higher $v\sin{i_\star}$ than the one derived from the RM analysis.

Nonetheless, the origin of the discrepancy between the two nights is rather unclear. The variable spot pattern on the stellar surface may cause this discrepancy. However, proving it or investigating it further is beyond the scope of this paper. Investigating activity indicators such as H$\alpha$ or Ca II H \& K lines would be an interesting possibility, but they are likely too noisy and do not contain enough information to trace the spot geometry in a way that would be useful to understand RM systematics.

\begin{figure}
\resizebox{\hsize}{!}{\includegraphics{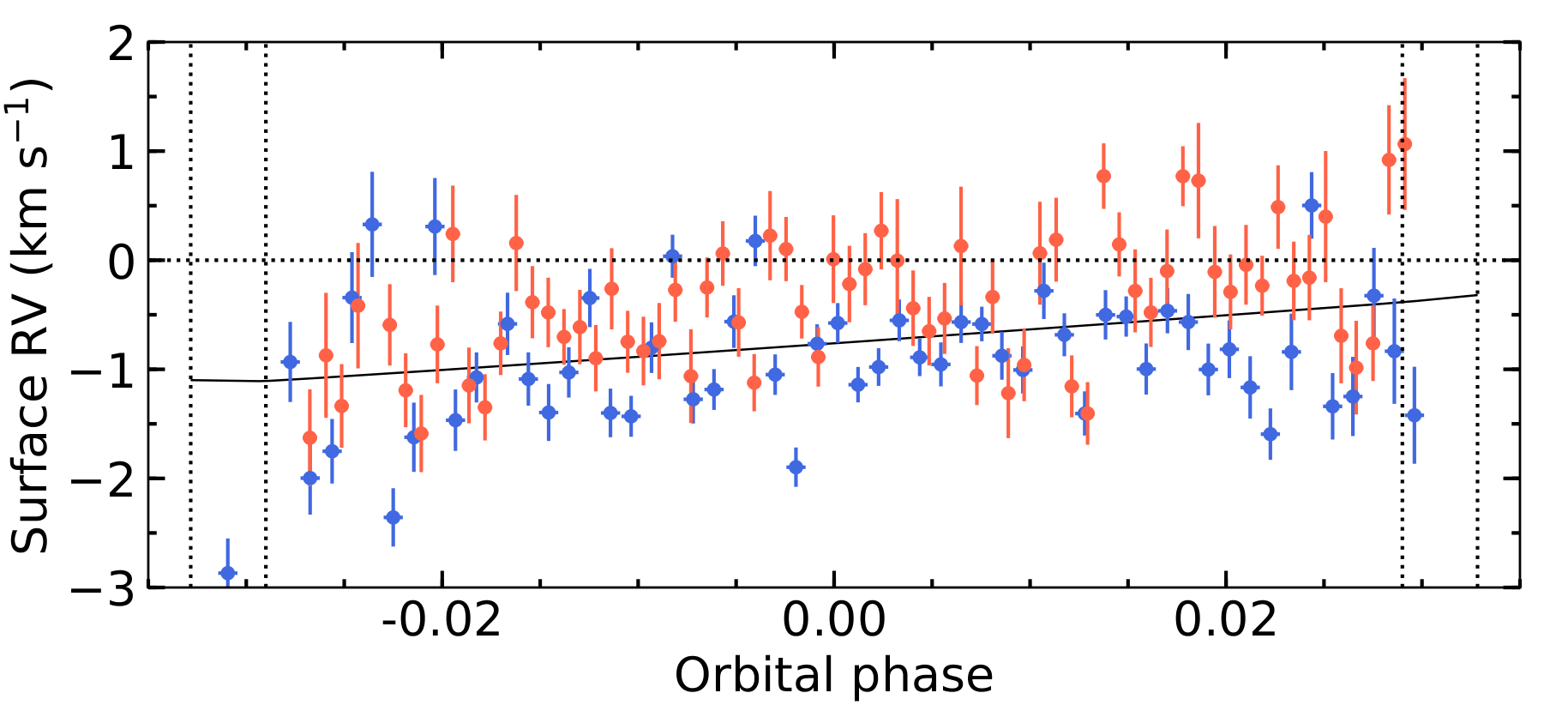}}
\caption{RVs of the stellar surface regions occulted by KELT-11\,b during the 16-02 (blue points) and 07-03 (orange points) nights. The solid black line is the best-fit model to both nights adjusted together. The vertical dotted lines indicate transit contacts.}
\label{fig:rvloc}
\end{figure}

As a result, we provide as well the results of the two nights fitted together. The best-fit model is shown in Figure \ref{fig:rvloc}. The derived values are $v\sin{i_\star}=1.99^{+0.06}_{-0.07}$ km s$^{-1}$ and $\lambda=-77.86^{+2.36}_{-2.26}{}^\circ$. The highly negative, projected obliquity value indicates a near polar orbit, as can be seen in Figure \ref{fig:rvloc} where most of the measured surface RVs are negative. 

The near polar orbit of KELT-11\,b may appear as an unexpected result. Indeed, previous statistical analyses have shown that hot Jupiters around cool stars ($T_\mathrm{eff}$ < 6250 K) tend to live on aligned orbits (e.g., \citet{Winn2010, Albrecht2012, Albrecht2022}). This stellar temperature threshold, also dubbed the Kraft break \citep{Kraft1967}, is linked to the transition to F8 stars, for which the convective zone is too thin to efficiently realign their systems via tidal interactions. KELT-11, with an effective temperature of $T_\mathrm{eff}$ = 5375 $\pm$ 25 K, should have then damped the obliquity of its hot Jupiter companion because of its extended convective region.

We compute a characteristic realignment time scale $\tau$, which represents the typical time needed for an efficient angular momentum realignment between the stellar spin-axis and the planetary orbit. Using the formula provided in \citet[][Eq.~(3)]{Hansen2012} and the parameters of Table  \ref{table:params}, we find $\tau$ $\cong$ 70 Gyr. This time scale being way higher than the age of the Universe, KELT-11 is actually unable to realign the orbit of its hot Jupiter, in agreement with our result. This is primarily due to the very strong dependence of this time scale on the semi-major axis. The separation of KELT-11\,b (a = 0.06230 $\pm$ 0.00104 AU) is in reality substantially higher than the fiducial separation of planets expected to be efficiently realigned (0.02 AU, \citet{Hansen2012}). Our results highlight the fact that the Kraft break should be taken with caution, as other important parameters should be accounted for. The case of KELT-11\,b, combined with the previous results reported in the literature, suggests that hot Jupiters may be born with random orientations only to be realigned by tides if possible.

\section{Modeling of the sodium absorption with MERC}
\label{sect:merc}
\begin{figure}[htb]
\resizebox{\hsize}{!}{\includegraphics[trim= 0cm 4cm 0cm 4cm]{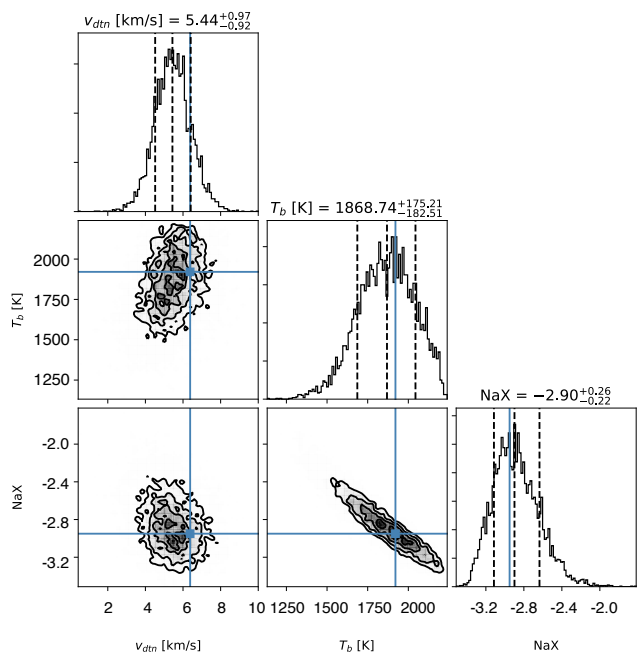}}
\caption{Posterior distribution of isothermal line retrieval with an added day-to-night side wind throughout the atmosphere. This is the best fit model for the here presented dataset and is also shown in Figure \ref{fig:wind_isodtn} in Appendix \ref{appendix:winds}.}
\label{fig:wind_isodtn_best}
\end{figure}

We applied the atmospheric retrieval code MERC to the detected sodium doublet. MERC combines a quasi 3D atmospheric wind model with a multinested-sampling retrieval algorithm to distinguish different wind patterns in the intermediate atmosphere. For the basics of MERC see \citet{Seidel2020a}, in this work, MERC is applied in its upgraded form, including solid body atmospheric winds and planetary rotation as described in \citet{Seidel2021}. The different wind patterns are isothermal, super-rotational ($\mathrm{srot}_{\cos\theta}$), day-to-night side ($\mathrm{dtn}_{\cos\theta}$), and a radial, outward (vertical) wind ($\mathrm{ver}_{\cos\theta}$). We note that our model assumes local thermodynamic equilibrium, which may underestimate the temperature probed \citep{Fisher2019}.

With the multinested-sampling approach, the different models are ranked compared to a basic model via the Bayesian evidence (see \citet{Seidel2020a} for more details). We use an isothermal model with no additional winds as the basic model, which only makes use of the planetary rotation and thermal broadening for the line fitting. Compared to this base model, any wind provides a better fit, while a day-to-night side wind at $5.44 \kms$ as the best fit outperforming all other applied scenarios (see Fig. \ref{fig:wind_isodtn_best}). The ranking via the Bayesian evidence can be interpreted using the Jeffrey scale \citep[see also][]{Skilling2006, Trotta2008}. The applied prior ranges for each model can be found in the Appendix \ref{appendix:winds} in Table \ref{table:wind_priors}. Each run was started with 5000 live points and convergence took on average 12\,000 steps. The computing cost for these models is relatively high since they are quasi-3D and all posteriors have converged with each best fit within 1$\sigma$ indicated in blue (Appendix \ref{appendix:winds}).

\begin{table}
\caption{Comparison of the different models.}
\label{table:comparison}
\centering
\begin{tabular}{l c c l}
\hline
\hline
Model & $\ln\mathcal{Z}$   &  $\ln\mathcal{B}_{01}$ & Strength of evidence  \tablefootmark{a}  \\
\hline
isothermal  &  $555.74\pm0.24$   & -  & - \\
$\mathrm{ver}_{\cos\theta}$  &  $558.00\pm0.25$   & $2.26$  & Weak - moderate \\
$\mathrm{srot}_{\cos\theta}$  &  $559.81\pm0.35$   & $4.07$  & Moderate\\
$\mathrm{dtn}_{\cos\theta}$  &  $562.87\pm0.33$   & $7.13$  & Strong \\
\hline
\end{tabular}
\tablefoot{\tablefoottext{a}{The base model to calculate $\ln\mathcal{B}_{01}$ is the isothermal model without added wind patterns. The comparison stems from the Jeffrey scale. }}
\end{table}

\begin{figure*}[htb!]
\resizebox{\hsize}{!}{\includegraphics[angle=90,trim= 6.2cm 0cm 6.5cm 0cm]{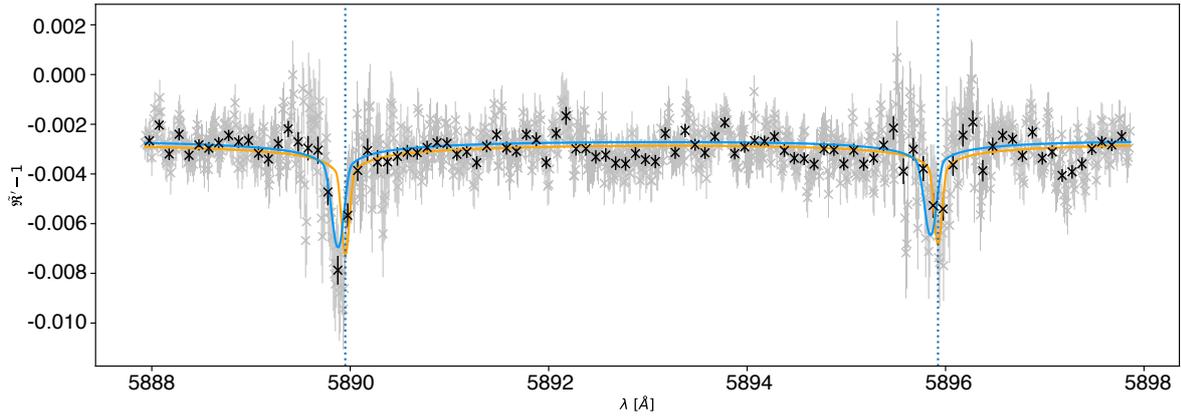}}
\caption{Sodium doublet of KELT-11\,b with two best fits retrieved with MERC overlaid. In gray, the original data in absorption, in black, the data binned by x5 for better visibility. In orange the best fit applying the basic model with no winds, only planetary rotation is shown, in blue the overall best fit is shown, for the model with a day-to-night side wind at approximately $5 \kms$.}
\label{fig:wind_full_fit}
\end{figure*}

\begin{figure}[htb]
\resizebox{\hsize}{!}{\includegraphics[trim= 0cm 8cm 0cm 8cm]{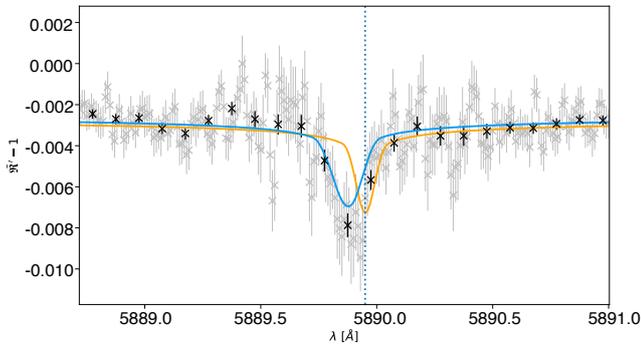}}
\caption{Zoom-in of Figure \ref{fig:wind_full_fit} on the D2 sodium line. See Figure \ref{fig:wind_full_fit} for further information.}
\label{fig:wind_zoom_fit}
\end{figure}

The fit of the best fit model of the day-to-night side wind can be seen in blue in the zoom-in of the D2 sodium line, Fig. \ref{fig:wind_zoom_fit} and Fig. \ref{fig:wind_full_fit} showing the full wavelength range of the sodium doublet for completeness. For comparison purposes, the basic model with no winds is shown in orange. Due to the increased noise in the D1 line, the fit is mainly driven by the deeper D2 line, which is shown in the zoom-in.

Two other worlds were studied so far with the use of MERC: HD\,189\,733\,b \citet{Seidel2020a} and WASP-76\,b \citet{Seidel2021}, two Jupiter-sized worlds at differing distances from their host stars, orbiting respectively a K-type star with a temperature of $\sim 4900~$K and an F-type star with a temperature of $\sim 6300~$K \citet{Bouchy2005,West2016}. 
On both planets, large sodium features were found, probing high up into their respective atmospheres. While both planets vary widely not only in their stellar parameters but also in their mass-radius ratio, both exhibit a strong, radial, vertical wind in their upper atmospheres transporting sodium upward to the thermosphere. This scenario is further corroborated by the large atmospheric temperatures retrieved with MERC which are a resulting overestimation due to the change of sodium density generated by the strong wind patterns \citet{Seidel2021}.

Here, however, for the most bloated planet in the sample of three, KELT-11\,b, completely different atmospheric conditions were encountered. The sodium feature probes fewer atmospheric layers and despite its bloated nature and comparable closeness to its host star with HD\,189\,733\,b as well as similar temperatures between KELT-11 and HD\,189\,733, and thus similar irradiation, KELT-11\,b does not show significant line broadening.

We theorise that this could have two likely causes: firstly, KELT-11\,b, in theory, shows similar vertical winds to the other two mentioned planets that are driven by the atmospheric material escaping the gravitational pull, but sodium is trapped in the lower layers of the atmosphere. This coincides with our best fit model showing day-to-night side winds, similar to WASP-76b's lower atmosphere.

Or secondly, the driver behind the strong vertical atmospheric winds in WASP-76\,b and HD\,189\,733\,b lies elsewhere and is not linked to the apparent bloating of the planetary atmosphere. A possibility is that the ionised sodium, which recombined further up to neutral sodium is dragged from the day-to-night side wind or super-rotational stream along the magnetic field lines, a theory explored in \citet{Seidel2020a} based on work in \citet{Cauley2019}.

The curious divergence of the inflated sub-Saturn KELT-11\,b from other inflated and highly irradiated exoplanets with a sodium signature shows not only the need for careful analysis of the shape of existing sodium detections with cutting-edge high-resolution spectrographs, but also the need for atmospheric circulation models to include the intermediate atmosphere and the impact of magnetically driven atmospheric movement.

% CONCLUSIONS
\section{Conclusion}
We detect sodium in the upper layers of the atmosphere of KELT-11\,b with a relative absorption of 0.50 \% and 0.28 \% at 8$\sigma$ and 6$\sigma$, respectively.
The 3-day observation method is a promising observation technique for transit spectroscopy that allows for the study of planets with longer transits. While more demanding in telescope time, it can provide high-precision results on types of exoplanets not well represented in studies yet, especially from the ground.
Our sodium detection is in accordance with the known features of other inflated sub-Saturns, of which thus far there are not enough candidates to search for characteristic trends.
Our analysis of the RM effect shows a low projected rotation velocity of 1.99$^{+0.06}_{-0.07}$ km s$^{-1}$ and spin-orbit angle of -77.86$^{+2.36}_{-2.26}{}^\circ$. Our simulation shows that the influence of the RM effect on the final transmission spectrum is negligible (see Appendix \ref{appendix:RM}). A better observation method for RM analysis would be to separate a transit into two nights with half a transit each and enough spectra pretransit and post-transit.
The modeling of the wind patterns from the sodium absorption with MERC results in day-to-night side winds at 5.4 $\pm$ 0.9 km s$^{-1}$ as best fit with no vertical winds, which is peculiar considering KELT-11\,b's extreme bloatedness.
Our results open the way to complementary atmospheric studies with other instruments such as NIRPS, ESPRESSO, and CRIRES. 
Inflated sub-Saturns have not been studied in detail yet, but their characteristics make them ideal targets for future observations.

%Acknowledgments - PlanetS
\begin{acknowledgements}
We acknowledge the Geneva exoplanet atmospheres group for their helpful feedback. We thank the referee for insightful comments that helped improve the manuscript. This work has been carried out within the framework of the National Centre of Competence in Research PlanetS supported by the Swiss National Science Foundation under grants 51NF40\textunderscore182901 and 51NF40\textunderscore205606. The authors acknowledge the financial support of the SNSF. This project has received funding from the European Research Council (ERC) under the European Union's Horizon 2020 research and innovation programme (project {\sc Four Aces}, grant agreement No. 724427; project {\sc Spice Dune}, grant agreement No. 947634). R.A. is a Trottier Postdoctoral Fellow and acknowledges support from the Trottier Family Foundation. This work was supported in part through a grant from FRQNT. M.L. acknowledges support of the Swiss National Science Fundation under grant number PCEFP2\textunderscore194576.
\end{acknowledgements}

% BIBLIOGRAPHY
\bibliographystyle{aa}    % style aa.bst
\bibliography{biblio}
%\printbibliography %Prints bibliography

% APPENDIX
\begin{appendix}

\appendix
\onecolumn
\newpage

% DACE fit
\section{DACE priors and fit}
\label{appendix:DACE}
%\LEt{ Single-sentence paragraphs are not allowed.\ Please expand here and below if possible.\ Also, you might consider rephrasing (e.g., writing something along the lines of "Readers can refer to...for...") to avoid directly addressing the reader.}See Sect. \ref{subsection:dace}, Table \ref{table:a1} and Fig. \ref{fig:DACE}.

\begin{figure*}[h]
\resizebox{\hsize}{!}{\includegraphics{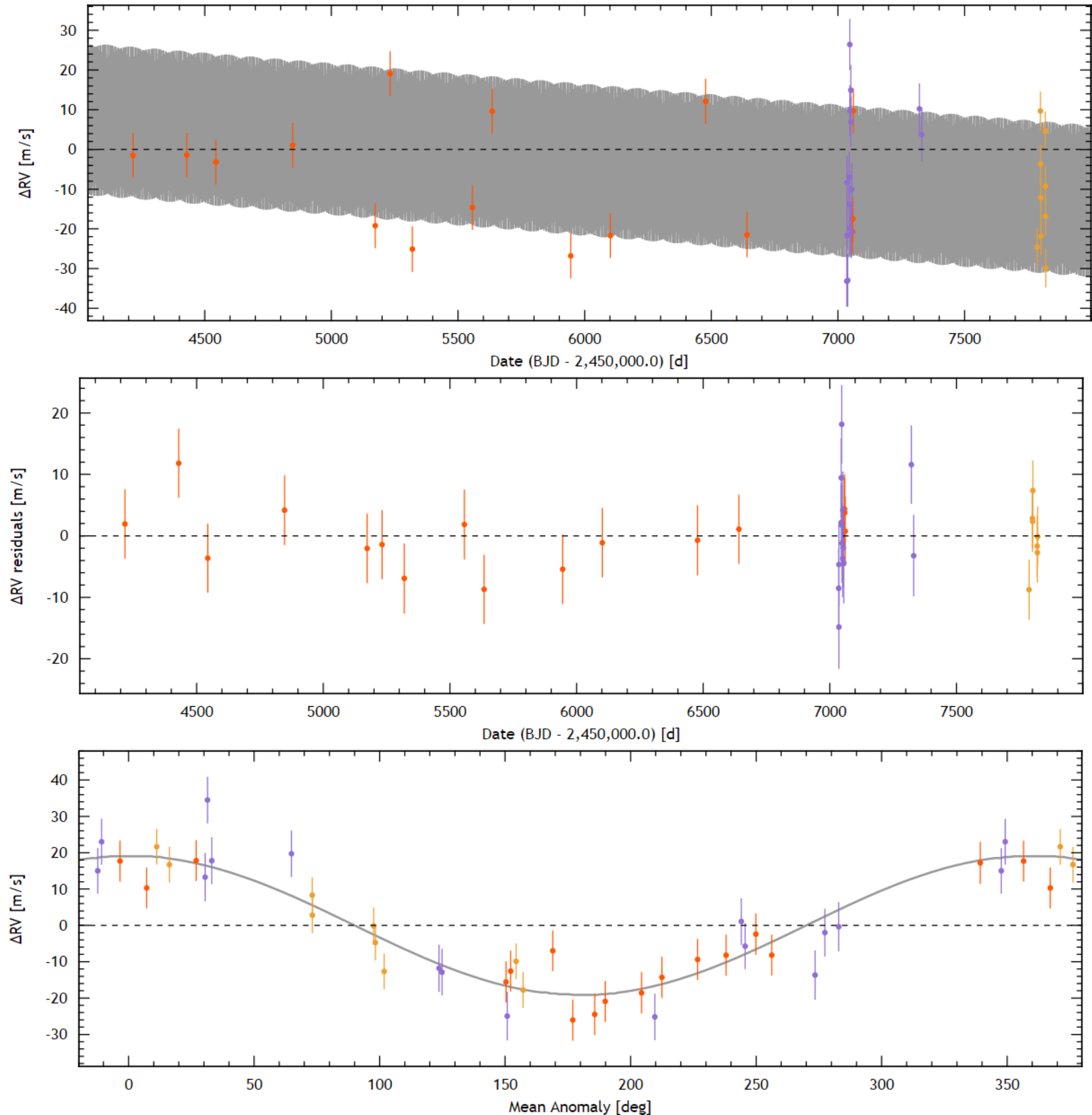}}
\caption{Keplerian orbit fit on DACE using available radial velocity measurements of KELT-11. \textit{Orange}: HIRES observations from \citet{Pepp2017}. \textit{Purple}: APF observations from \citet{Pepp2017}. \textit{Yellow}:  HARPS observations from this work (binned every 5 hours). \textbf{Top}: Time series signal with Keplerian fit in gray. \textbf{Middle}: Residuals from the Keplerian fit. \textbf{Bottom}: Phase-folded signal, compatible with a circular orbit.}
\label{fig:DACE}
\end{figure*}

\begin{table*}[h]
\large
\caption{Parameters used as priors for DACE refinement of orbital parameters of KELT-11 system}
\begin{threeparttable}
\begin{tabular}{l l}
\toprule \toprule
Parameter & Value [Unit] \\
\midrule
Stellar mass $M_\mathrm{\star}$  & 1.44 $\pm$ 0.07 $M_\mathrm{\odot}$ \tablefootmark{a} \\
Stellar radius $R_\mathrm{\star}$ & 2.69 $\pm$ 0.04 $R_\mathrm{\odot}$ \tablefootmark{a}\\
Parallax Par & 10.056 $\pm$ 0.053 mas \tablefootmark{b} \\
Orbital period P & 4.7362085 $\pm$ 0.0000038 d \tablefootmark{c}  \\
Transit Epoch $T_\mathrm{C}$ & 2457483.43043 $\pm$ 0.00080 $BJD_\mathrm{TDB}$ \tablefootmark{c} \\
Eccentricity e & 0 (fixed) \\
APF noise & 3 m/s (fixed) \\
HIRES noise & 2.5 m/s (fixed) \\
HARPS noise & 0.75 m/s (fixed) \\
\bottomrule
\end{tabular}
\begin{tablenotes}
\item[]References:
\item[a] \citet{Beatty2017}.
\item[b] \textit{Gaia} Archive \citet{Gaia2016, Gaia2018b}.
\item[c] \citet{Colon2020}.
\item[] \textbf{Notes}: Additional fitted parameters: Stellar jitter, linear drift, stellar RV semi-amplitude, stellar semi-major axis, planetary semi-major axis. %See Fig. \ref{fig:stats}
\end{tablenotes}
\end{threeparttable}
\label{table:a1}
\end{table*}

\newpage
\clearpage
% Molecfit
\section{Molecfit parameters}
\label{appendix:mol}
%See Sect. \ref{sect:molecfit}, Table \ref{table:mol1} and Table \ref{table:mol2}.

\begin{table*}[h]
\caption{Initial parameters for \textsc{Molecfit}, which were identical for every night.}
\label{table:mol1}
\begin{tabular}{l l l}
\toprule \toprule
Parameter & Value & Significance \\
\midrule
ftol & 10$^{-5}$ & $\chi^2$ convergence criterion \\
xtol & 10$^{-5}$ & Parameter convergence criterion \\
molecules & H$_2$O, O$_2$ \\
$n_\mathrm{cont}$ & 2 & Degree of coefficient for continuum fit \\
$a_\mathrm{0}$ & 13700 & Constant term for continuum fit \\
$n_\mathrm{\lambda}$ & 2 & Polynomial degree of the refined wavelength solution \\
$b_\mathrm{0}$ & 0 & Constant term for wavelength correction \\
$\omega_\mathrm{gaussian}$ & 4.5 & FWHM of Gaussian in pixels \\
kernel size & 3 \\
%pixel scale & 0.16 arcsec \\ Doesn't appear anymore in ver 4
slit width & 0.4 arcsec \\
MIPAS profile & equ & Equatorial profile \\
Atmospheric profile & 0 & Natural grid \\
PWV & -1 & Input for water vapor profile (-1 implies no scaling) \\
\bottomrule
\end{tabular}
\end{table*}

\begin{table*}[ht]
\caption{Fitted wavelength ranges with Molecfit for all nights}
\label{table:mol2}
\begin{tabular}{c}
\toprule
Fitted regions [$\mu$m] \\
\midrule
\multicolumn{1}{c}{-- \textit{H$_2$O regions} --} \\
0.591944 - 0.592046\\
0.592107 - 0.592264\\
0.593325 - 0.593570\\
0.594700 - 0.594937\\
\multicolumn{1}{c}{-- \textit{O$_2$ regions ($\gamma$ band)} --} \\
0.627752 - 0.628155\\
0.628348 - 0.628425\\
0.628545 - 0.628702\\
0.629101 - 0.629256\\
0.629384 - 0.629496\\
0.629664 - 0.629842\\
\multicolumn{1}{c}{-- \textit{H$_2$O regions} --} \\
0.647473 - 0.647740\\
0.648130 - 0.648309\\
0.648800 - 0.649337\\
0.651344 - 0.651750\\
0.652300 - 0.652805\\
0.654230 - 0.654750\\
\bottomrule
\end{tabular}
\end{table*}

\newpage
\clearpage
\section{Rossiter-McLaughlin simulation}

\begin{figure*}[h]
\resizebox{\hsize}{!}{\includegraphics{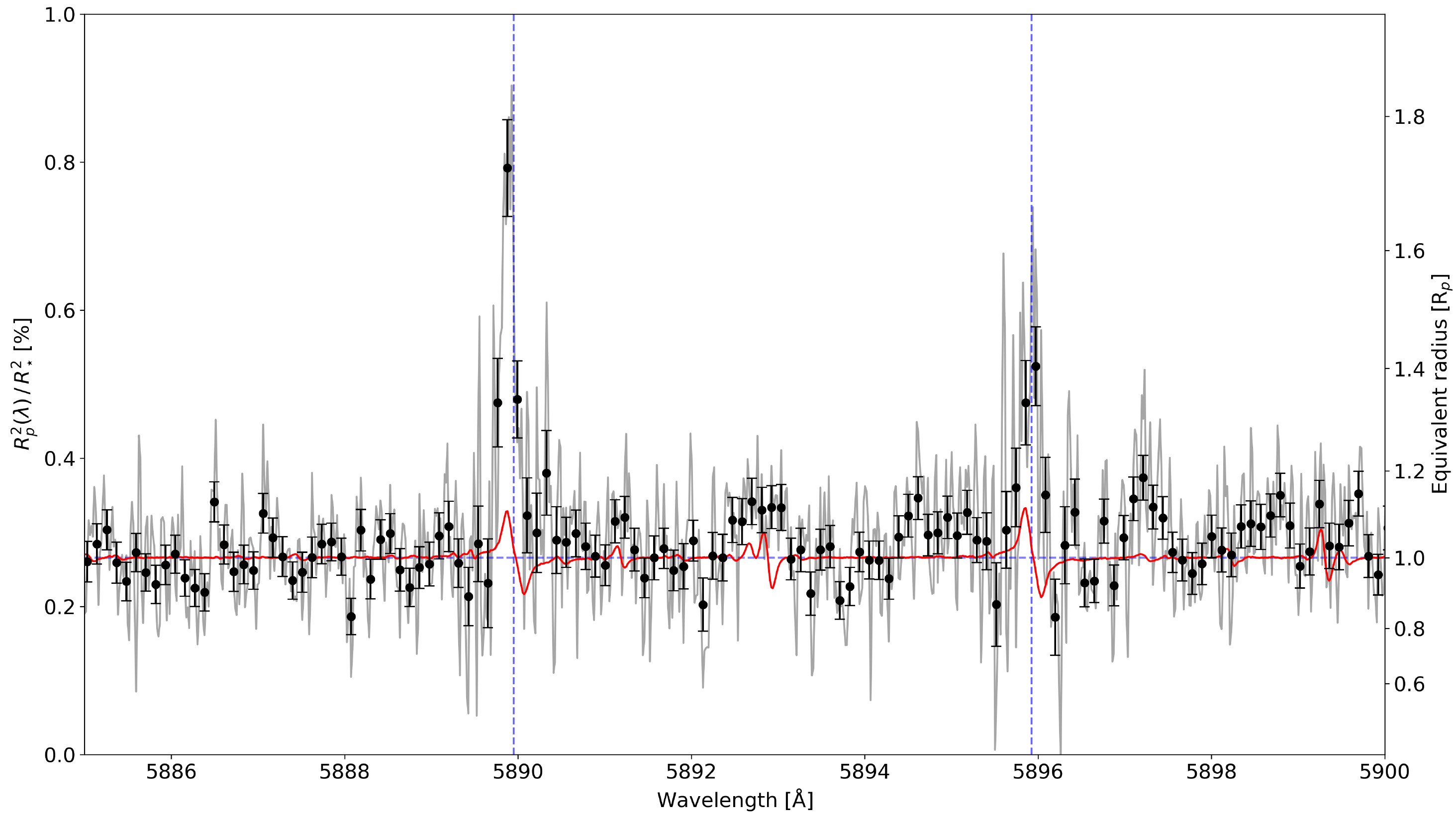}}
\caption{Simulation of the Rossiter-McLaughlin effect (RM, in red) compared to the final transmission spectrum, the same as obtained in Fig \ref{fig:finalts}.}
\label{fig:RM_sim}
\end{figure*}

\label{appendix:RM}
%See Sect \ref{section:RM}.

To verify the influence of the Rossiter-McLaughlin effect on the transmission spectrum (see Fig. \ref{fig:RM_sim}), we simulated the RM effect based on the master-out spectrum of Epoch 2. We generated mock in-transit spectra by subtracting a scaled-down and shifted version of the master-out to remove the spectrum occulted by the planet. The RV shift corresponds to the RVs computed on Figure \ref{fig:rvloc} for the corresponding orbital phases of all our spectra. These synthetic in-transit spectra are then plugged in the same pipeline, resulting in the red simulated transmission spectrum. This spectrum represents the impact of the pure RM effect for a planet with no atmosphere. We can see that the RM amplitude is similar to the noise level of the merged transmission spectrum, and thus we can neglect the impact of the RM effect in the current study.

\section{MERC retrieval prior and posterior distributions}

\label{appendix:winds}
%See Sect. \ref{sect:merc}, Table \ref{table:wind_priors}, and Figures \ref{fig:wind_isonone}, \ref{fig:wind_isover}, \ref{fig:wind_isosrot}, and \ref{fig:wind_isodtn}.

\begin{table*}[!h]
\caption{Overview of the different prior ranges of the models. For more information on the priors, see \citet{Seidel2020a}.}
\label{table:wind_priors}
\centering
\begin{tabular}{l c c c c c}
\hline
\hline
Model & $T_{\mathrm{iso}}$ [K]  &  NaX & $v_{\mathrm{srot}}$ [\kms] & $v_{\mathrm{dtn}}$ [\kms] & $v_{\mathrm{ver}}$ [\kms]     \\
\hline
isothermal  & [1000, 3000]    & [-4.0, -1.0]  & -  & -  &  - \\
$\mathrm{dtn}_{\cos\theta}$  &  [1000, 3000]   & [-4.0, -1.0]  & -  &  [0.1,25.0]  & -\\
$\mathrm{srot}_{\cos\theta}$  &  [1000, 3000]   & [-4.0, -1.0]  & [0.1,25.0]   &  - & - \\
$\mathrm{ver}_{\cos\theta}$ &  [1000, 3000]   & [-4.0, -1.0]  &  - & -  & [0.1,40.0] \\
\hline
\end{tabular}
\end{table*}

\begin{figure}[!h]
\begin{minipage}{0.48\textwidth}
\resizebox{\hsize}{!}{\includegraphics{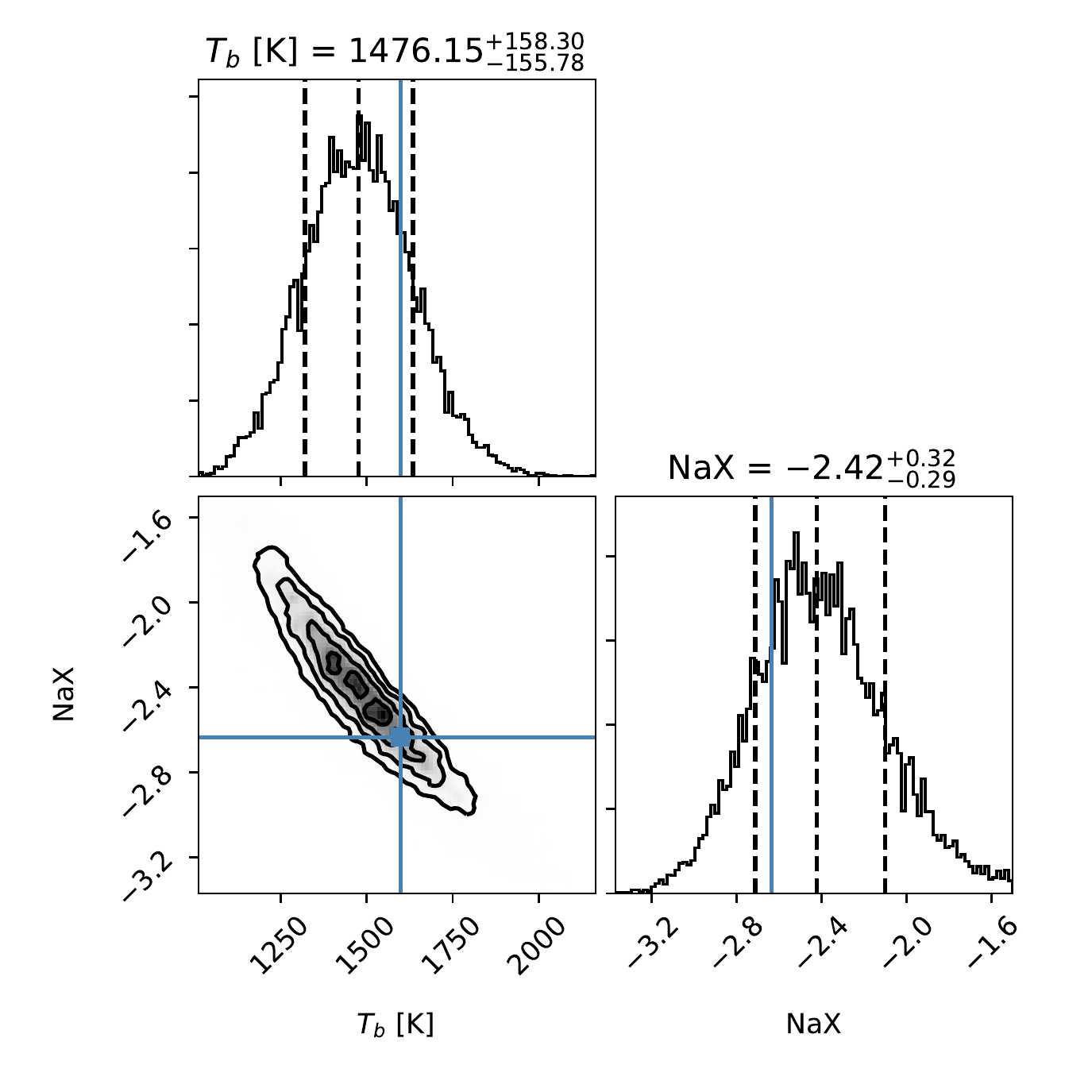}}
 \caption{Posterior distribution of isothermal line retrieval with no additional winds. This model is the basic model to which all other scenarios are compared. MERC includes planetary rotation and latitudinal wind variations due to solid body rotation in all models, including the basic model with no atmospheric winds.}
 \label{fig:wind_isonone}
%\end{figure}
\end{minipage}\hfill
\begin{minipage}{0.48\textwidth}
%\begin{figure}[!h]
\resizebox{\hsize}{!}{\includegraphics{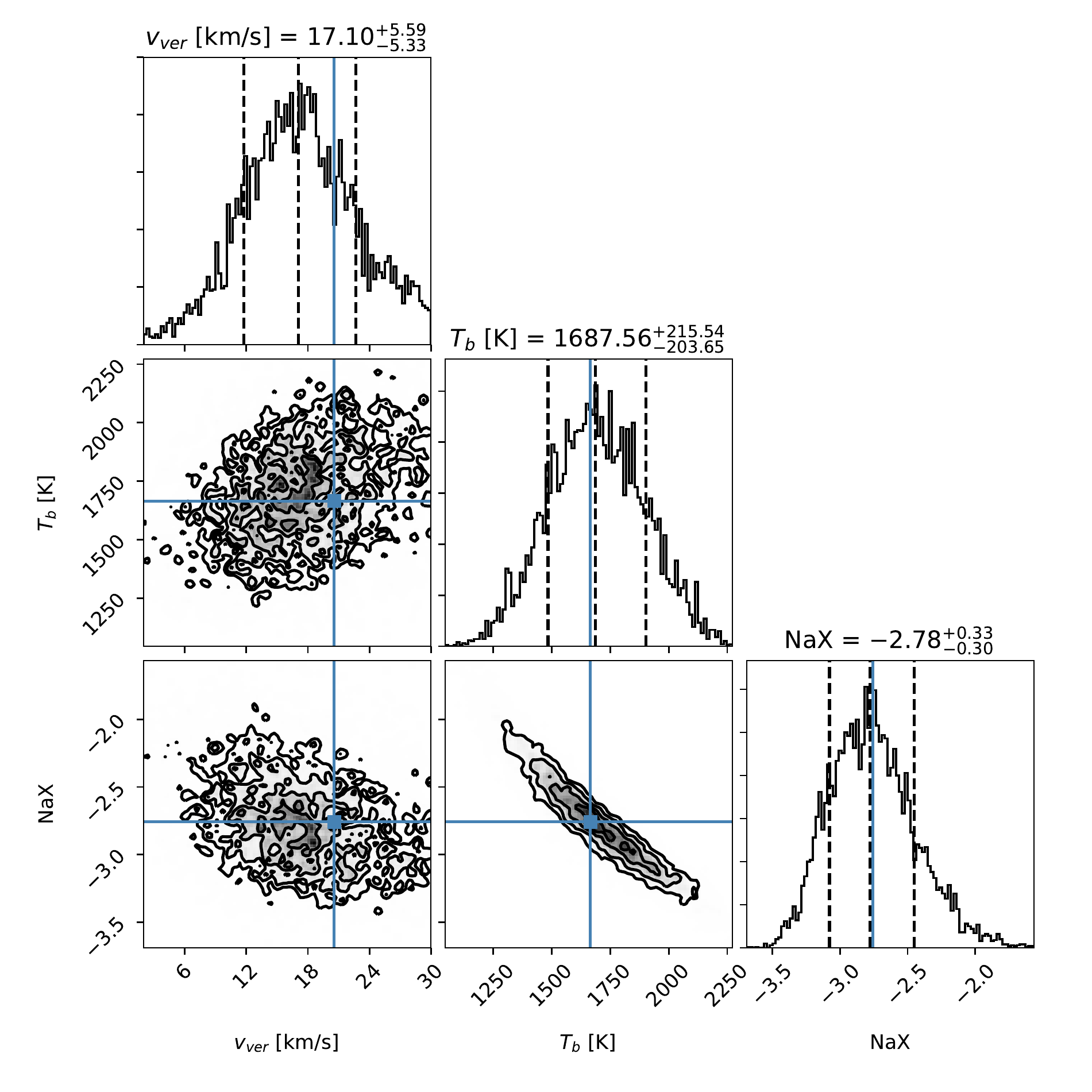}}
 \caption{Posterior distribution of isothermal line retrieval with an added vertical wind throughout the atmosphere.}
 \label{fig:wind_isover}
\end{minipage}
\end{figure}

\begin{figure}[!h]
\begin{minipage}{0.48\textwidth}
\resizebox{\hsize}{!}{\includegraphics{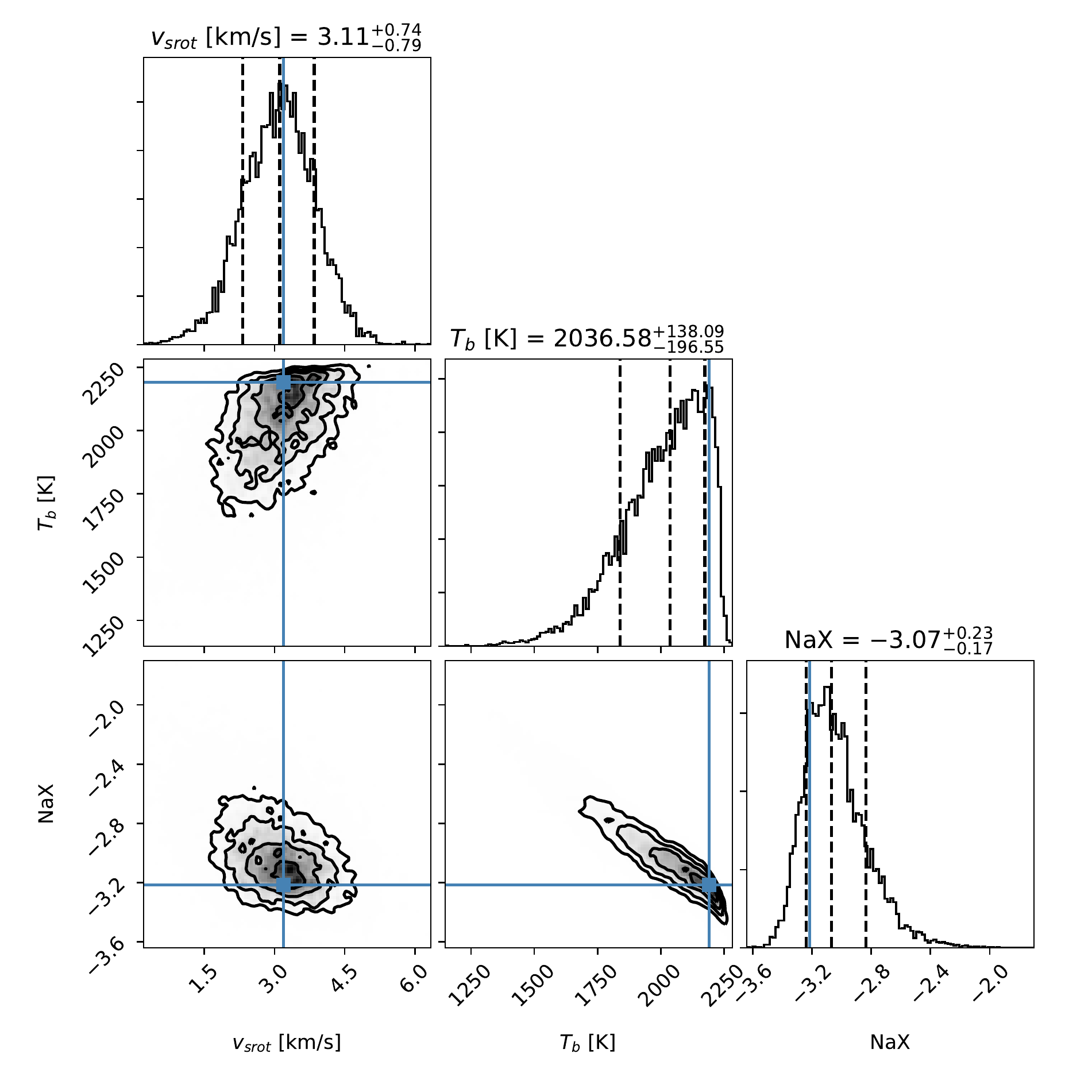}}
 \caption{Posterior distribution of isothermal line retrieval with an added super-rotational wind throughout the atmosphere.}
 \label{fig:wind_isosrot}
%\end{figure}
\end{minipage}\hfill
\begin{minipage}{0.48\textwidth}
%\begin{figure}[!h]
\resizebox{\hsize}{!}{\includegraphics{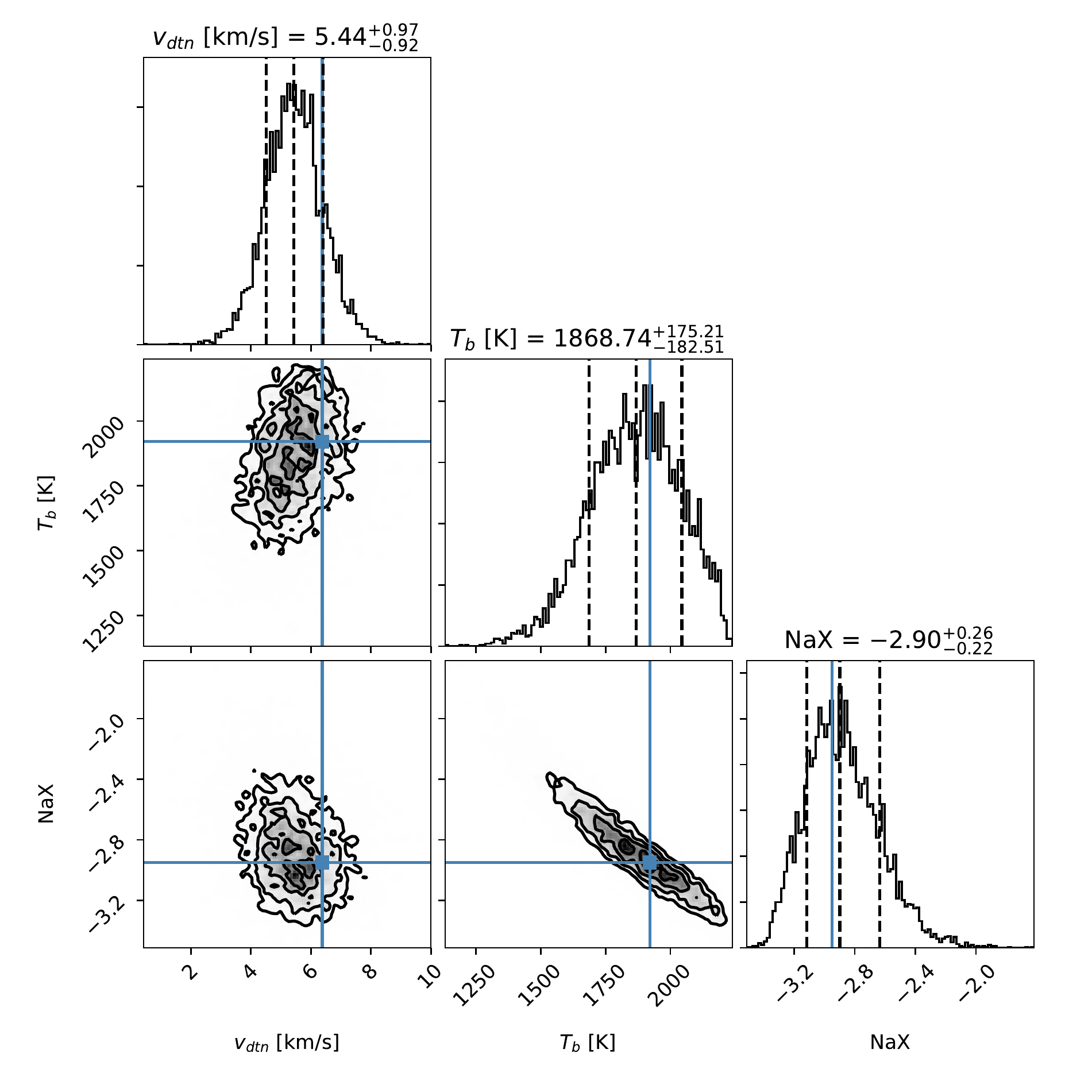}}
 \caption{Posterior distribution of isothermal line retrieval with an added day-to-night side wind throughout the atmosphere.}
 \label{fig:wind_isodtn}
 \end{minipage}
\end{figure}

\end{appendix}

\end{document}